\documentclass[twoside,11pt]{article}

\usepackage[preprint]{jmlr2e}

\usepackage{bm}
\usepackage{amsmath}
\usepackage{algorithm, algorithmic}

\usepackage{times}

\usepackage{subfigure}
\usepackage{multirow}

\def\c{{\bm c}}

\def\z{{\bm z}}

\jmlrheading{XX}{20XX}{XX-XX}{4/24; Revised XX/XX}{XX/XX}{XX-XX}{Author One and Author Two}
\ShortHeadings{TNPM for Directed and Bipartite Networks}{Jing, Li, Wang and Wang}

\title{Two-way Node Popularity Model for Directed and Bipartite Networks}
\author{\name Bing-Yi Jing \email jingby@sustech.edu.cn \\
       \addr Department of Statistics and Data Science\\
       Southern University of Science and Technology\\
       Shenzhen {\rm518055}, China
       \AND
       \name Ting Li \email tingeric.li@polyu.edu.hk \\
       \addr Department of Applied Mathematics\\
       The Hong Kong Polytechnic University\\
       Hong Kong 999077, China
       \AND
       \name Jiangzhou Wang \email wangjz695@szu.edu.cn \\
       \addr School of Mathematical Sciences, Institute of Statistical Sciences\\
       Shenzhen University\\
       Shenzhen {\rm 518060}, China
       \AND
       \name Ya Wang  \email wangya@sustech.edu.cn \\
       \addr Department of Statistics and Data Science\\
       Southern University of Science and Technology\\
       Shenzhen {\rm518055}, China
       }

\editor{}

\begin{document}
\maketitle

\begin{abstract}
There has been extensive research on community detection in directed and bipartite networks. However, these studies often fail to consider the popularity of nodes in different communities, which is a common phenomenon in real-world networks. To address this issue, we propose a new probabilistic framework called the Two-Way Node Popularity Model (TNPM). The TNPM also accommodates edges from different distributions within a general sub-Gaussian family. We introduce the Delete-One-Method (DOM) for model fitting and community structure identification, and provide a comprehensive theoretical analysis with novel technical skills dealing with sub-Gaussian generalization. Additionally, we propose the Two-Stage Divided Cosine Algorithm (TSDC) to handle large-scale networks more efficiently. Our proposed methods offer multi-folded advantages in terms of estimation accuracy and computational efficiency, as demonstrated through extensive numerical studies. We apply our methods to two real-world applications, uncovering interesting findings.
\end{abstract}

\vspace{2mm}


\section{Introduction}\label{sec:intro}

Community detection is a valuable tool for understanding the structure of a network and has been applied in various fields, including biology (\cite{calderer2021community, li2021super}), social science (\cite{wu2020deep,jing2022community}), and global trading analysis (\cite{jing2021community}). While several models have been proposed for community detection in undirected networks, the study of community detection in directed and bipartite networks is relatively limited.

One reason is that directed networks are more complex than undirected ones as they involve both outgoing and incoming links. Therefore, traditional definitions of clustering problems, such as intra-cluster and inter-cluster edge density, cannot be extended directly to directed networks (\cite{zhang2021directed}). Bipartite networks, on the other hand, have nodes divided into two sets and edges that only connect nodes from different sets. This feature violates the assumption of symmetric relationships between nodes in undirected scenarios. 

To address this gap, several studies have been conducted. The pseudo-likelihood approach has been used to identify out- and in-community structures in \cite{amini2013pseudo} and \cite{wang2023fast}. \cite{rohe2016co} proposes the Stochastic co-Blockmodel (ScBM) and its extension, the Degree Corrected ScBM (DC-ScBM), which considers degree heterogeneity to model directed networks. \cite{wang2020spectral} analyzes the theoretical guarantees for the algorithm D-SCORE (\cite{ji2016coauthorship}) and its variants designed under DC-ScBM. \cite{zhou2019analysis} studies spectral clustering algorithms designed by a data-driven regularization of the adjacency matrix under ScBM. In \cite{zhang2021directed}, authors embed nodes with concentration restrictions to help identify communities.

However, all of the above methods overlook the heterogeneous popularities of nodes across different communities. Such structure has been widely observed and discussed in undirected networks (\cite{sengupta2018block,noroozi2021estimation}). Addressing this, \cite{sengupta2018block} proposes the Popularity Adjusted Stochastic Block Model (PABM), which models the edge probability between two nodes as a product of node popularity parameters. PABM provides a flexible way of modeling the probability of connections and allows nodes in the same community to exhibit heterogeneous popularities across different communities.  

\begin{figure}
\centering
\includegraphics[trim=0 8mm 0 0,scale=0.8]{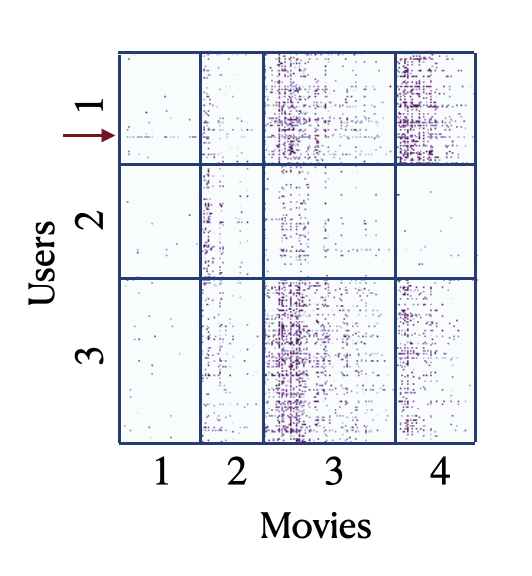}
\caption{The adjacency matrix of MovieLens 100K Data set is rearranged by the clustering results obtained through the proposed TSDC method, with blue lines marking the cluster boundaries.}
\label{WTD Figure}
\end{figure}

Diverse node popularity patterns are also widely present in directed or bipartite scenarios. One motivating example is the MovieLens 100K data set (\cite{harper2015movielens}), which depicts a bipartite network where entries represent user-to-movie ratings. Illustrated in Figure \ref{WTD Figure}, the adjacency matrix is organized according to clustering results, with cluster boundaries marked by blue lines. This visualization reveals that group 1 users predominantly prefer movies in categories 3 and 4, though there's limited but notable interest in categories 1 and 2, with a few exceptions (highlighted by a red arrow) showing curiosity in category 1. Such diversity in node popularities reveals various consumer patterns and contributes to understanding the different behaviors of users with respect to movies from different categories.

Existing algorithms for undirected networks are not easily extendable to directed and bipartite scenarios. For example, the algorithm proposed in \cite{sengupta2018block} is limited to networks with a small number of communities (less than 3). On the other hand, while the Sparse Subspace Clustering (SSC) algorithm proposed in \cite{noroozi2021estimation} can handle large-scale networks, it is sensitive to noise, as demonstrated by the simulation results in Section \ref{section:simulation}. Moreover, our numerical studies indicate that naively applying these methods to directed and bipartite networks often leads to poor performance.

In this paper, we introduce the Two-Way Node Popularity Model (TNPM), a comprehensive probabilistic framework designed to model directed and bipartite networks with community structures and node popularities. Moreover, the TNPM allows each link to be generated from different distributions within the sub-Gaussian family. The new model presents significant challenges in model fitting due to the use of two distinct sets of scaling parameters to characterize node popularities for the out- and in-communities separately. Our main contributions are listed as follows:
\begin{itemize}
    \item We propose the Two-Way Node Popularity Model (TNPM) to model directed and bipartite networks with community structures and node popularities. The model also generalizes link distributions to the sub-Gaussian family.
    \item To fit the model, we introduce the Delete-One-Method (DOM) with theoretical guarantees, and the Two-Stage Divided Cosine Algorithm (TSDC) for large-scale networks. Both methods have been empirically proven to be superior to state-of-the-art methods.
    \item We prove the consistency of the DOM under the TNPM, including the consistency of the estimated probability matrix and the detected out- and in-community structures. We adopt a new strategy to directly upper bound the operator norm of random matrices to overcome the technical issue raised by following former works, which need to prove the concentration inequality for the Lipschitz function of independent sub-Gaussian random variables.
\end{itemize}

To the best of our knowledge, this is the first systematic study of directed and bipartite networks that considers node popularities, and the study on sub-Gaussian generalization might be of independent interest.

The rest of the paper is organized as follows. Section \ref{section:model} introduces the Two-Way Node Popularity Model (TNPM). In Section \ref{section:method}, we propose the Delete-One-Method (DOM) and the Two-Stage Divided Cosine Algorithm (TSDC). We explore the theoretical properties in Section \ref{section:theory}. Section \ref{section:simulation} and \ref{section:real} present extensive simulations and real data applications to demonstrate the advantages of the proposed methods.

\section{Two-way Node Popularity Model}\label{section:model}

Consider a bipartite network $\mathcal{G} (U,V,E)$ with two sets of nodes $U$ and $V$ indexed as $1, \ldots, n$ and $1, \ldots, m$, respectively. The directed network can be viewed as a special case of the bipartite network when $U=V$. Therefore, from this point onward, we will solely focus on bipartite graphs, and all the algorithms and conclusions derived will also be applicable to directed networks.

Let $A=\left(A_{ij}\right)_{i,j=1}^{n,m}$ denote the adjacency matrix of the network, where $A_{i,j}$ represents the weights from node $i$ in set $U$ to node $j$ in set $V$. We use $A_{i \cdot}$ to denote the $i$-th row of matrix $A$, and $A_{\cdot j}$ to denote the $j$-th column of $A$. The community structure associated with nodes in set $U$ is referred to as the out-community, and the community structure associated with nodes in set $V$ is referred to as the in-community. Let $K$ and $L$ denote the number of out-communities and in-communities, respectively. The distinct blocks are denoted as $\mathcal{N}_k$ and $\mathcal{M}_l$ for all $k=1,\ldots,K$ and $l=1,\ldots,L$.

For brevity, we introduce the following notations. For any set $\Omega$, denote cardinality of $\Omega$ by $|\Omega|$. For any numbers $a$ and $b, a \wedge b=\min (a, b)$ and $\lfloor a \rfloor$ represents the largest integer less than or equal to $a$. For any integer $I$, denote $[I]$ as the set $\{1, 2, \ldots, I\}$. We define $n_k=\left|\mathcal{N}_k\right|$ and $m_l=\left|\mathcal{M}_l\right|$ for all $k \in [K]$ and $l \in [L]$. Furthermore, let $\bm c \in \left\{1, 2, \ldots, K\right\}^{n} \triangleq [K]^n$ and $\bm z \in \left\{1, 2, \ldots, L\right\}^{m} \triangleq [L]^m$ represent the vectors of out-community and in-community assignments, respectively. Specifically, $c_i=k$ if and only if node $i$ belongs to out-community $\mathcal{N}_k$, and $z_j=l$ if and only if node $j$ belongs to in-community $\mathcal{M}_l$. Denote $\mathcal{M}_{n, K}$ and  $\mathcal{M}_{m, L}$ as the collections of clustering matrices $C \in\{0,1\}^{n \times K}$ and $Z \in\{0,1\}^{m \times L}$, respectively, where $C_{ik}=1$ if $i \in \mathcal{N}_k$ and $Z_{jl}=1$ if $j \in \mathcal{M}_l$.

To model the popularity of nodes in rows and columns separately, we propose a Two-Way Node Popularity Model (TNPM):
\begin{equation}
\label{eqn:(1)}
P_{ij}=\mathbb{E}\left(A_{ij}\right)=\Lambda_{i\bm z_j} \widetilde{\Lambda}_{j\bm c_i},
\end{equation}
where $\Lambda_{il}$, $1\leq i \leq n$, $1\leq l \leq L$, and $\widetilde{\Lambda}_{jk}$, $1\leq j \leq m$, $1\leq k \leq K$ are the node popularity parameters. $\Lambda_{il}$ describes the popularity of node $i$ among the in-community ${\mathcal{M}}_l$, while $\widetilde{\Lambda}_{jk}$ represents the popularity of node $j$ among the out-community $\mathcal{N}_k$.

Additionally, given $\bm c$ and $\bm z$, the $A_{ij}$'s are assumed to be mutually independent and follow a sub-Gaussian distribution with 
$\mathbb{E}(A_{ij})=P_{ij}$, such as Bernoulli, Binomial and Normal  distribution. It is crucial for modeling real data, since there is no prior knowledge of specific distributions.

The mean structure of the adjacency matrix $A$ under TNPM exhibits a block rank-one structure, which constitutes the main idea of constructing the algorithms for fitting TNPM. Specifically, let's consider a rearranged version $P(\bm c, \bm z)$ of the matrix $P$, where the rows and columns are allocated based on community membership. For instance, nodes belonging to out (in) -community 1 occupy the first $n_1$ ($m_1$) rows (columns), nodes from out (in) -community 2 occupy the subsequent $n_2$ ($m_2$) rows (columns), and so on. We denote the $\left(k,l\right)$-th block of matrix $P(\bm c, \bm z)$ as $P^{\left(k,l\right)}(\bm c, \bm z)$. The sub-matrix $P^{\left(k,l\right)}(\bm c, \bm z) \in \mathbb{R}^{n_k \times {m}_l}$ corresponds to the nodes in the community pair $\left(\mathcal{N}_k,{\mathcal{M}}_l\right)$, respectively. Furthermore, we have $P_{ij}^{k,l}= \Lambda_{i_k l} \widetilde{\Lambda}_{j_l k}$, where $i_k$ represents the $i$-th element in $\mathcal{N}_k$ and $j_l$ represents the $j$-th element in ${\mathcal{M}}_l$. As a result, the matrices $P^{(k,l)}(\bm c, \bm z)$ are rank-one matrices with unique singular vectors. In fact, we can express them as
\begin{equation}
\label{eqn:(2)}
P^{(k,l)}(\bm c, \bm z)=V^{(k,l)}\left[\widetilde{V}^{(l,k)}\right]^T,
\end{equation}
where $V^{(k,l)}$ and $\widetilde{V}^{(l,k)}$  are vectors with elements $V^{(k,l)}_i=\Lambda_{i_k l}$, for $i \in [n_k]$ and $i_k \in \mathcal{N}_k,$ and $\widetilde{V}^{(l,k)}_j=\widetilde{\Lambda}_{j_l k}$, for $j \in [{m}_l]$ and $j_l \in {\mathcal{M}}_l$.
Therefore, we can rewrite $P(\bm c, \bm z)$ as
\begin{equation}
\label{eqn:(3)}
P(\bm c, \bm z)=\left[\begin{array}{cccc}
V^{(1,1)}\left(\widetilde{V}^{(1,1)}\right)^{T} & V^{(1,2)}\left(\widetilde{V}^{(2,1)}\right)^{T} & \ldots & V^{(1, L)}\left(\widetilde{V}^{(L, 1)}\right)^{T} \\
V^{(2,1)}\left(\widetilde{V}^{(1,2)}\right)^{T} & V^{(2,2)}\left(\widetilde{V}^{(2,2)}\right)^{T} & \ldots & V^{(2, L)}\left(\widetilde{V}^{(L, 2)}\right)^{T} \\
\vdots & \vdots & \ldots & \vdots \\
V^{(K, 1)}\left(\widetilde{V}^{(1, K)}\right)^{T} & V^{(K, 2)}\left(\widetilde{V}^{(2, K)}\right)^{T} & \ldots & V^{(K, L)}\left(\widetilde{V}^{(L, K)}\right)^{T}
\end{array}\right].
\end{equation}

Rank-one structures have also been observed in undirected networks \citep{ noroozi2021sparse, noroozi2021estimation}. However, it remains a significant challenge to identify these structures without prior knowledge of the node memberships, especially when dealing with large K or/and L values. Direct application of previous methods can result in poor fitting performance and can also be time-consuming, as demonstrated in our empirical studies.

\section{Methodology}\label{section:method}

In this section, we introduce the Delete-One-Method (DOM) and the Two-Stage Divided Cosine Algorithm (TSDC) for the purpose of model fitting and community detection under the TNPM.

\subsection{The Delete-One-Method (DOM)}
With the observation of the block rank-one structure shown in equation \eqref{eqn:(3)}, we propose the objective function defined as follows:
\begin{equation}
\label{eqn:(10)}
 Loss(\bm c, \bm z)=\sum_{k=1}^{K}\sum_{l=1}^{L}\left \|A^{(k,l)}(\bm c, \bm z)-V^{(k,l)}\left[\widetilde{V}^{(l,k)}\right]^T\right \|_F^2,
\end{equation}
where $A^{(k,l)}(\bm c, \bm z)$ represents the $\left(k,l\right)$-th block of $A\left(\bm c, \bm z \right)$, and $A\left(\bm c, \bm z \right)$ is the rearranged matrix of $A$ according to $\bm c$ and $\bm z$.

To address the identifiability issue in the recovery of $V^{(k,l)}$ and $\widetilde{V}^{(l,k)}$, we introduce $\Theta^{(k,l)}$ as the notation for $V^{(k,l)}\left[\widetilde{V}^{(l,k)}\right]^T$ and focus on recovering the uniquely defined rank-one matrix $\Theta^{(k,l)}$. In addition, since the numbers of communities $K$ and $L$ are usually unknown, following the idea in \cite{noroozi2021sparse, noroozi2021estimation}, a penalty on $K$ and $L$ is introduced to safeguard against choosing too many communities. Consequently, our optimization problem can be formulated as follows:

\begin{equation}
\label{eqn:(11)}
\left(\widehat{\Theta}, \hat{\bm c}, \hat{\bm z}, \widehat{K}, \widehat{L}\right)=\arg \min_{\bm c, \bm z, K, L}
\left\{\sum_{k=1}^{K}\sum_{l=1}^{L}\left \|A^{(k,l)}(\bm c, \bm z)-\Theta^{(k,l)}\right \|_F^2+Pen(n,m,K,L)\right\}\\
\end{equation}
$\qquad \qquad \qquad \qquad \qquad \qquad \text{s.t.} \operatorname{rank}\left(\Theta^{(k, l)}\right)=1 ; \quad k \in [K] ;\quad l \in [L]$,\\
where $\widehat{\Theta}$ is the block matrix with blocks $\widehat{\Theta}^{(k,l)}$, and $Pen(n, m, K, L)$ is the item of penalty and will be defined later.

If $\hat{\bm c}$, $\hat{\bm z}$, $\widehat{K}$, and $\widehat{L}$ were known, the optimal solution to problem \eqref{eqn:(11)} would be obtained by the rank-one approximations $\widehat{\Theta}^{(k, l)}$ of the sub-matrix $A^{(k,l)}(\hat{\bm c}, \hat{\bm z})$. These approximations can be expressed as
\begin{equation}
\label{eqn:(12)}
\hat{\Theta}^{k,l}(\hat{\bm c},\hat{\bm z})= \Pi_{\hat{u},\hat{v}}	(A^{(k,l)}(\hat{\bm c}, \hat{\bm z}))=\hat{\sigma}_1^{(k,l)}\hat{u}^{(k,l)}(\hat{\bm c},\hat{\bm z})(\hat{v}^{(k,l)}(\hat{\bm c},\hat{\bm z}))^T
\end{equation}
where $\hat{\sigma}_1^{(k,l)}$ represents the largest singular value of $A^{(k,l)}(\hat{\bm c}, \hat{\bm z})$, and $\hat{u}^{(k,l)}(\hat{\bm c},\hat{\bm z})$ and $\hat{v}^{(k,l)}(\hat{\bm c},\hat{\bm z})$ are the corresponding singular vectors. Here, the operation $\Pi(\cdot)$ denotes the rank-one projection.
Plugging \eqref{eqn:(12)} into \eqref{eqn:(11)}, the optimization problem \eqref{eqn:(11)} could be rewritten as
\begin{equation}
\label{eqn:(13)}
\left(\hat{\bm c}, \hat{\bm z}, \widehat{K}, \widehat{L}\right)=\arg \min_{\bm c, \bm z, K, L}\left\{\sum_{k=1}^{K}\sum_{l=1}^{L}\left \|A_{(\mathcal{N}_k , {\mathcal{M}}_l)}^{(\bm c, \bm z)}-\widehat{\Theta}^{(k,l)}\right \|_F^2+Pen(n, m, K, L)\right\}.
\end{equation}
In order to obtain $(\hat{\bm c},\hat{\bm z})$, one need to solve optimization problem \eqref{eqn:(13)} for every $K$ and $L$, obtaining
\begin{equation}
\label{eqn:(cz)}
\left(\hat{\bm c}_K,\hat{\bm z}_L\right)=\arg \min_{ \genfrac{}{}{0pt}{}{\bm z \in {[L]}^m}{\bm c \in {[K]}^n} } \sum_{k=1}^{K}\sum_{l=1}^{L}\left \|A_{(\mathcal{N}_k , {\mathcal{M}}_l)}^{(\bm c, \bm z)}-\prod_{\hat{u},\hat{v}}(A_{(\mathcal{N}_k , {\mathcal{M}}_l)}^{(\bm c, \bm z)})\right \|_F^2,
\end{equation}
and then find $\hat{K}$ and $\hat{L}$ as
\begin{equation}
(\hat{K},\hat{L})= \arg \min_{K,L}
\left\{\sum_{k=1}^{K}\sum_{l=1}^{L}\left \|A_{(\mathcal{N}_k , {\mathcal{M}}_l)}^{(\hat{\bm c}_K,\hat{\bm z}_L)}-\prod_{\hat{u},\hat{v}}(A_{(\mathcal{N}_k , {\mathcal{M}}_l)}^{(\hat{\bm c}_K,\hat{\bm z}_L)})\right \|_F^2+Pen(n, m, K, L)\right\}.
\end{equation}

The optimization problem \eqref{eqn:(cz)} constitutes the most crucial part of the fitting algorithm, and its optimization is often NP-hard. Consequently, the development of efficient algorithms for approximating the optimization of \eqref{eqn:(cz)} is of utmost significance. To address this challenge, we introduce an alternating update algorithm and integrate a Delete-One-Method (DOM) within the iteration process, resulting in a substantial reduction in computational complexity. 

Specifically, at the t-th step, when  $\left(\bm c^{(t)}, \bm z^{(t)}\right)$ are given, we update $\bm c$ and $\bm z$ separately to obtain $\bm c^{(t+1)}$ and $\bm z^{(t+1)}$. To provide a detailed explanation, given $\bm c^{(t)}$ and $\bm z^{(t)}$, the sub-optimization task related to $\bm c^{(t+1)}$ can be expressed as
\begin{equation}
\bm c^{(t+1)}=\arg \min_{\bm c \in {[K]}^n} \sum_{k=1}^{K}\sum_{l=1}^{L}\left \|A_{(\mathcal{N}_k , {\mathcal{M}}_l)}^{(\bm c, \bm z^{(t)})}-\prod_{\hat{u},\hat{v}}(A_{(\mathcal{N}_k , {\mathcal{M}}_l)}^{(\bm c, \bm z^{(t)})})\right \|_F^2.
\end{equation}
In particular, for each $i$ $\in$ $[n]$, we have:
\begin{equation}
\label{eqn:(17)}
\bm c_{i}^{(t+1)}=\arg \min_{\widetilde{\bm c}_{i}^{(t)} \in {[K]}} \sum_{k=1}^{K}\sum_{l=1}^{L}\left \|A_{(\mathcal{N}_k , {\mathcal{M}}_l)}^{(\widetilde{\bm c}^{(t)}, \bm z^{(t)})}-\prod_{\hat{u},\hat{v}}(A_{(\mathcal{N}_k , {\mathcal{M}}_l)}^{((\widetilde{\bm c}^{(t)}, \bm z^{(t)})})\right \|_F^2,
\end{equation}
where $\widetilde{\bm c}_{p}^{(t)}=\bm c_{p}^{(t)}$ when $p \ne i $.

The minimization problem \eqref{eqn:(17)} is computationally expensive since it requires calculating the Frobenius norm error for each $K \times L$ block. To simplify this calculation, we propose subtracting the value of the right side of \eqref{eqn:(17)} for $A_{-i}$, where $A_{-i}$ refers to the adjacency matrix $A$ with the $i$-th row deleted
\begin{equation}
\begin{aligned}
\label{eqn:(18)}
\bm c_{i}^{(t+1)}=\arg \min_{\widetilde{\bm c}_{i}^{(t)} \in [K]} \Bigg\{
\sum_{k=1}^{K}\sum_{l=1}^{L}&\left \|A_{(\mathcal{N}_k , {\mathcal{M}}_l)}^{(\widetilde{\bm c}^{(t)}, \bm z^{(t)})}-\prod_{\hat{u},\hat{v}}(A_{(\mathcal{N}_k , {\mathcal{M}}_l)}^{(\widetilde{\bm c}^{(t)}, \bm z^{(t)})})\right \|_F^2\\
&-\sum_{k=1}^{K}\sum_{l=1}^{L}\left \|A_{(\mathcal{N}_k\backslash{i}, {\mathcal{M}}_l)}^{(\widetilde{\bm c}^{(t)}, \bm z^{(t)})}-\prod_{\hat{u},\hat{v}}(A_{(\mathcal{N}_k\backslash{i} , {\mathcal{M}}_l)}^{(\widetilde{\bm c}^{(t)}, \bm z^{(t)})})\right \|_F^2
\Bigg\}.\\
\end{aligned}
\nonumber	
\end{equation}
Thus, we obtain the simplified expression as follows:
\begin{equation}
\begin{aligned}
\label{eqn:(18_1)}
\bm c_{i}^{(t+1)}=\arg \min_{\widetilde{\bm c}_{i}^{(t)} \in [K]} \Bigg\{
\sum_{l=1}^{L}&\left \|A_{(\mathcal{N}_{\widetilde{\bm c}_{i}^{(t)}} , {\mathcal{M}}_l)}^{(\widetilde{\bm c}^{(t)}, \bm z^{(t)})}-\prod_{\hat{u},\hat{v}}(A_{(\mathcal{N}_{\widetilde{\bm c}_{i}^{(t)}} , {\mathcal{M}}_l)}^{(\widetilde{\bm c}^{(t)}, \bm z^{(t)})})\right \|_F^2\\
&-\sum_{l=1}^{L}\left \|A_{(\mathcal{N}_{\widetilde{\bm c}_{i}^{(t)}}\backslash{i}, {\mathcal{M}}_l)}^{(\widetilde{\bm c}^{(t)}, \bm z^{(t)})}-\prod_{\hat{u},\hat{v}}(A_{(\mathcal{N}_{\widetilde{\bm c}_{i}^{(t)}}\backslash{i} , {\mathcal{M}}_l)}^{(\widetilde{\bm c}^{(t)}, \bm z^{(t)})})\right \|_F^2
\Bigg\}.\\
\end{aligned}
\end{equation}
Similarly, the sub-optimization task related to $\bm z^{(t+1)}$ can be described as follows:
\begin{equation}
\begin{aligned}
\bm z^{(t+1)}=\arg \min_{\bm z \in {[L]}^m} 
\Bigg\{\sum_{k=1}^{K}\sum_{l=1}^{L}& \left \|A_{(\mathcal{N}_k , {\mathcal{M}}_l)}^{(\bm c^{(t+1)}, \bm z)}-\prod_{\hat{u},\hat{v}}(A_{(\mathcal{N}_k , {\mathcal{M}}_l)}^{(\bm c^{(t+1)}, \bm z)})\right \|_F^2 \\
&-\sum_{k=1}^{K}\sum_{l=1}^{L}\left \|A_{(\mathcal{N}_{k} , {\mathcal{M}}_{\widetilde{\bm z}_{j}^{(t)}\backslash{j}})}^{\bm (c^{(t+1)}, \widetilde{\bm z}^{(t)})}-\prod_{\hat{u},\hat{v}}(A_{(\mathcal{N}_{k} , {\mathcal{M}}_{\widetilde{\bm z}_{j}^{(t)}\backslash{j}})}^{\bm (c^{(t+1)},\widetilde{\bm z}^{(t)})})\right \|_F^2\Bigg\}.
\nonumber 
\end{aligned}
\end{equation}
Consequently, for each $j$ $\in$ $[m]$, the sub-optimization task for $\bm z_j^{(t+1)}$ is defined as follows:
\begin{equation}
\begin{aligned}
\label{eqn:(19)}
\bm z_{j}^{(t+1)}=\arg \min_{\widetilde{\bm z}_{j}^{(t)} \in [L]} \Bigg\{
\sum_{k=1}^{K}&\left \|A_{(\mathcal{N}_{k} , {\mathcal{M}}_{\widetilde{\bm z}_{j}^{(t)}})}^{\bm (c^{(t+1)}, \widetilde{\bm z}^{(t)} )}-\prod_{\hat{u},\hat{v}}(A_{(\mathcal{N}_{k} , {\mathcal{M}}_{\widetilde{\bm z}_{j}^{(t)}})}^{(\bm c^{(t+1)},\widetilde{\bm z}^{(t)})})\right \|_F^2\\
&-\sum_{k=1}^{K}\left \|A_{(\mathcal{N}_{k} , {\mathcal{M}}_{\widetilde{\bm z}_{j}^{(t)}\backslash{j}})}^{(\bm c^{(t+1)}, \widetilde{\bm z}^{(t)})}-\prod_{\hat{u},\hat{v}}(A_{(\mathcal{N}_{k} , {\mathcal{M}}_{\widetilde{\bm z}_{j}^{(t)}\backslash{j}})}^{(\bm c^{(t+1)}, \widetilde{\bm z}^{(t)})})\right \|_F^2
\Bigg\}.\\
\end{aligned}
\end{equation}
The overall process of the DOM algorithm can be summarized as in Algorithm \ref{algo_DOM}.

\begin{algorithm}
	\renewcommand{\algorithmicrequire}{\textbf{Input:}}
	\renewcommand{\algorithmicensure}{\textbf{Output:}}
	\caption{Delete-One-Method Algorithm}
	\label{alg1}
	\begin{algorithmic}[1]
	   \REQUIRE observed adjacency matrix $A$, $\epsilon$, $iter_{\max}$
	   \STATE Initialization: $\left(\bm c^{(0)},\bm z^{(0)}\right)$, and set $t=0$.
	   \WHILE{$t < iter_{\max}$}
	   \FOR{$i \gets 1$ to $n$}
	  	   \STATE $\bm c^{(t+1)}_{i}$ is calculated by \eqref{eqn:(18)} with $\{{c_{1}^{(t)}, \ldots, c_{i-1}^{(t)}}\}$ replaced by $\{{c_{1}^{(t+1)}, \ldots, c_{i-1}^{(t+1)}}\}$.
	   \ENDFOR
	   \FOR{$j \gets 1$ to $m$}
	  	   \STATE $\bm z^{(t+1)}_{j}$ is calculated by \eqref{eqn:(19)} with $\{{z_{1}^{(t)}, \ldots, z_{j-1}^{(t)}}\}$ replaced by $\{{z_{1}^{(t+1)}, \ldots, z_{j-1}^{(t+1)}}\}$, and with $\bm c^{(t)}$ replaced by $\bm c^{(t+1)}$.
	   \ENDFOR
	   \IF{$\frac{\left|\bm L(\bm c^{(t+1)}, \bm z^{(t+1)})-\bm L(\bm c^{(t)}, \bm z^{(t)})\right|}{\bm L(\bm c^{(t)}, \bm z^{(t)})} < \epsilon $}
	   \STATE break
	   \ENDIF
	   \STATE $t\leftarrow t+1$
       \ENDWHILE
       \ENSURE return $\hat{\bm c}=\bm c^{(t)}$, $\hat{\bm z}=\bm z^{(t)}.$
	\end{algorithmic}
\label{algo_DOM}
\end{algorithm}

In the initialization step of Algorithm \ref{algo_DOM}, we employ a combination of SVD and K-means clustering to obtain $\bm c^{(0)}$ and $\bm z^{(0)}$. 
Given the adjacency matrix $A$, we initially apply SVD ($A\approx U\Sigma V^T$) to extract features from rows and columns. Here, $U \in \mathbb{R}^{n \times K}$ and $V \in \mathbb{R}^{m \times L}$ represent the row and column features, respectively. Subsequently, we employ K-means clustering on $U$ and $V$ to obtain the row cluster labels $\bm c^{(0)}$ and column cluster labels $\bm z^{(0)}$, respectively. Simulation studies have demonstrated that this initialization method yields more satisfactory results compared to other naive approaches.

\subsection{Two-Stage Divided Cosine Algorithm (TSDC)}

While the DOM algorithm successfully decreases computational complexity and can handle network data with thousands of nodes within an acceptable time range, it still has limitations when it comes to effectively deal with large-scale network data. Hence, to tackle this drawback, we propose a more computationally efficient algorithm called the Two-Stage Divided Cosine Algorithm (TSDC). The number of communities, denoted as $K$ and $L$, is assumed to be known throughout this subsection.

Let's consider a block $P^{(k,l)}(\bm c, \bm z)$, where out-node $i$ and out-node $j$ belong to the same community $\mathcal{N}_k$. In this case, corresponding to the $i_k$-th and $j_k$-th rows of $P^{(k,l)}(\bm c, \bm z)$, as given by the equation \eqref{eqn:(2)}, we have
\begin{equation}
P^{(k,l)}_{i_k .}(\bm c, \bm z)= V^{(k,l)}_{i_k}\left(\widetilde{V}^{(l,k)}\right)^{T},\\
P^{(k,l)}_{j_k .}(\bm c, \bm z)= V^{(k,l)}_{j_k}\left(\widetilde{V}^{(l,k)}\right)^{T}.	
\end{equation}
Thus, the cosine similarity between $P^{(k,l)}{i_k .}(\bm c, \bm z)$ and $P^{(k,l)}{j_k .}(\bm c, \bm z)$ is equal to 1. However, when out-node $i$ and out-node $j$ do not belong to the same community, the cosine similarity between $P^{(k,l)}{i_k .}(\bm c, \bm z)$ and $P^{(k,l)}_{j_k .}(\bm c, \bm z)$ is strictly less than 1 under the assumption of pairwise linear independence, i.e. $\widetilde{V}^{(l,k)}$ and $\widetilde{V}^{(l,k')}$ are linearly independent for any $k\neq k'$.

In view of this, we propose a similarity measure called Block Cosine Similarity to combine cosine similarities throughout all the column  communities. Let $A_{i}.$ and $A_{j.}$ denote the $i$-th and $j$-th rows of $A$, respectively. Given the column community label $\bm z$, the Block Cosine Similarity between $A_{i}.$ and $A_{j}.$ is defined as
\begin{equation}
BlockCos\left(A_{i}.,A_{j}.\right)= \sum_{l=1}^{L}cos(A_{i \mathcal{M}_l},A_{j \mathcal{M}_l}).\nonumber
\end{equation}
Similarly, for any two in-nodes $i$ and $j$, given the row community label $\bm c$, the Block Cosin Similarity between $A._{i}$ and $A._{j}$ is defined as:
\begin{equation}
BlockCos\left(A._{i},A._{j}\right)= \sum_{l=1}^{K}cos(A_{ \mathcal{N}_k i},A_{\mathcal{N}_k j}).\nonumber
\end{equation}

Basing on the Block Cosin Similarity, we propose a two-stage algorithm to facilitate the community detection for both rows and columns. The first stage aims to detect the row assignment $\bm c$ given  $\bm z$, and the corresponding objective function is defined as
\begin{equation}
	\bm L(\bm c | \bm z)=\sum_{i=1}^{n}BlockCos(A_{i \cdot},\bm \mu_{\bm c_i}.)/L. \nonumber
\end{equation}
Here, $\bm \mu \in \mathbb{R}^{K\times m}$ represents the community centers for rows, and the similarity function used is the $BlockCos$. 
Similarly, in the second stage, we update $\bm z$ given $\bm c$, and the objective function is defined as
\begin{equation}
	\widetilde{\bm L}(\bm z| \bm c)=\sum_{j=1}^{m}BlockCos(A_{\cdot j},\widetilde{\bm \mu}_{\cdot \bm z_j})/K, \nonumber
\end{equation}
where $\widetilde{\bm \mu} \in \mathbb{R}^{n\times L}$ represents the column community centers.

The proposed objective function is minimized by alternatively updating $\left(\bm c, \bm z\right)$ and $\left(\bm \mu, \widetilde{\bm \mu}\right)$. Specifically, given $\left(\bm c^{(t)}, \bm z^{(t)}\right)$ at the $t$-th step, we first obtain $\left(\bm \mu^{(t)}, \widetilde{\bm \mu}^{(t)}\right)$ through the suboptimization task
\begin{equation}
\bm \mu^{(t)}= \max_{\bm \mu}\bm L(\bm c^{(t)} | \bm z^{(t)})=\max_{\bm \mu}\sum_{i=1}^{n}BlockCos(A_{i}.,\bm \mu_{\bm c_i^{(t)}\cdot})/L,\nonumber
\end{equation}
\begin{equation}
\widetilde{\bm \mu}^{(t)}=\max_{\widetilde{\bm \mu}}\widetilde{\bm L}(\bm z^{(t)}| \bm c^{(t)})=\max_{\widetilde{\bm \mu}}\sum_{j=1}^{m}BlockCos(A_{\cdot j},\widetilde{\bm \mu}_{\cdot \bm z_j^{(t)}})/K.	\nonumber
\end{equation}

Given a block $P^{(k,l)}(\bm c, \bm z)$, its row cluster center, $\mu_{k,\widetilde{\mathcal{N}}_l}\in \mathbb{R}^{\tilde{n}_l}$, and column cluster center, $\tilde{\mu}_{\mathcal{N}_k,l}\in \mathbb{R}^{n_k}$, can be derived as
\begin{equation}
\mu_{k,\widetilde{\mathcal{N}}_l}=\frac{\mu_{k,\widetilde{\mathcal{N}}_l}^\prime}{\left\|\mu_{k,\widetilde{\mathcal{N}}_l}^\prime\right\|_1}\times \tilde{n}_l,\quad \text{where}\quad
\mu_{k,\widetilde{\mathcal{N}}_l}^\prime=\frac{\sum_{i=1}^{n_k}\frac{P^{(k,l)}_{{i}.}(\bm c, \bm z)}{\left\|P^{(k,l)}_{{i}.}(\bm c, \bm z)\right\|_2}}{n_k},\label{muupdate}
\end{equation}
\begin{equation}
\tilde{\mu}_{\mathcal{N}_k,l}=\frac{\tilde{\mu}_{\mathcal{N}_k,l}^\prime}{\left\|\tilde{\mu}_{\mathcal{N}_k,l}^\prime\right\|_1}\times n_k,\quad \text{where}\quad
\tilde{\mu}_{\mathcal{N}_k,l}^\prime=\frac{\sum_{j=1}^{\tilde{n}_l}\frac{P^{(k,l)}_{.{j}}(\bm c, \bm z)}{\left\|P^{(k,l)}_{.{j}}(\bm c, \bm z)\right\|_2}}{\tilde{n}_l}.\label{mutildeupdate}
\end{equation}
Combining the cluster centers for each block, we obtain
\begin{eqnarray}
\bm \mu=\left(
\begin{array}{cccc}
\mu_{11} & \mu_{12} & \ldots  & \mu_{1m}\\
\mu_{21} & \mu_{22} & \ldots  & \mu_{2m}\\
\vdots & \vdots & &\vdots \\
\mu_{K1} & \mu_{K2} & \ldots  & \mu_{Km}\\
\end{array}
\right) \in \mathbb{R}^{K*m}, \;
\widetilde{\bm \mu}=
\left(
\begin{array}{cccc}
\widetilde{\mu}_{11} & \widetilde{\mu}_{12} & \ldots  & \widetilde{\mu}_{1L}\\
\widetilde{\mu}_{21} & \widetilde{\mu}_{22} & \ldots  & \widetilde{\mu}_{2L}\\
\vdots & \vdots & &\vdots \\
\widetilde{\mu}_{n1} & \widetilde{\mu}_{n2} & \ldots  & \widetilde{\mu}_{nL}\\
\end{array}
\right)\in \mathbb{R}^{n*L}. \nonumber
\end{eqnarray}

We update $\bm c$ and $\bm z$ separately to obtain $\bm c^{(t+1)}$ and $\bm z^{(t+1)}$. Given $\bm c^{(t)}$, $\bm z^{(t)}$, $\bm \mu^{(t)}$, $\widetilde{\bm \mu}^{(t)}$, the optimization task related to $\bm c^{(t+1)}$ becomes
\begin{equation}
\max_{\bm c}\bm L(\bm c | \bm z^{(t)})=\max_{\bm c}\sum_{i=1}^{n}BlockCos(A_{i}.,\bm \mu^{(t)}_{\bm c_i}.)/L.
\nonumber 	
\end{equation}
Specially, for each $i$ $\in$ $[n]$,
\begin{equation}
\bm c^{(t+1)}_{i}=\mathop{\arg\max}_{1\leq \bm c_i \leq K}BlockCos(A_{i}.,\bm \mu^{(t)}_{\bm c_i}.)/L.
\nonumber	
\end{equation}
Similarly, the sub-optimization task related to $\bm z^{(t+1)}$ is
\begin{equation}
\max_{\bm z}\widetilde{\bm L}(\bm z |\bm c^{(t+1)})=\max_{\bm z}\sum_{j=1}^{m}BlockCos(A_{\cdot j},\widetilde{\bm \mu}^{(t)}_{\cdot\bm z_j})/K.
\nonumber 	
\end{equation}
Thus, for each $j$ $\in$ $[m]$,
\begin{equation}
\bm z^{(t+1)}_{i}=\mathop{\arg\max}_{1\leq \bm z_j \leq L}BlockCos(A_{\cdot j},\widetilde{\bm \mu}^{(t)}_{\cdot\bm z_j})/K.
\nonumber	
\end{equation}
Next, given $\bm c^{(t+1)}$ and $\bm z^{(t+1)}$, we update $\bm \mu$ and $\widetilde{\bm \mu}$ using equations (\ref{muupdate}) and (\ref{mutildeupdate}), respectively. The whole algorithm can be summarized as shown in Algorithm \ref{algo_TSDC}. The initialization of$\left(\bm c^{(0)}, \bm z^{(0)}\right)$  follows the same procedure as in the DOM method.
\begin{algorithm}
	\renewcommand{\algorithmicrequire}{\textbf{Input:}}
	\renewcommand{\algorithmicensure}{\textbf{Output:}}
	\caption{Two-stage Divided Cosine Algorithm}
	\label{alg2}
	\begin{algorithmic}[1]
	   \REQUIRE observed adjacency matrix $A$, $\epsilon$, $iter_{\max}$
	   \STATE Initialization: $\left(\bm c^{(0)},\bm z^{(0)}\right)$, and set $t=0$.
	   \WHILE{$t < iter_{\max}$}
	   \STATE $\bm \mu^{(t)}$ and $\widetilde{\bm \mu}^{(t)}$ are calculated according to (\ref{muupdate}) and (\ref{mutildeupdate}), respectively.
	   \STATE $\bm c^{(t+1)}_{i}=\mathop{\arg\max}_{1\leq \bm c_i \leq K}BlockCos(A_{i\cdot},\bm \mu^{(t)}_{\bm c_i \cdot})
$, $1\leq i \leq n$.
       \STATE $\bm z^{(t+1)}_{i}=\mathop{\arg\max}_{1\leq \bm z_j \leq L}BlockCos(A_{\cdot j},\widetilde{\bm \mu}^{(t)}_{\cdot\bm z_j})$,$1\leq j \leq m$.
       \IF{$\frac{\left|\bm L(\bm c^{(t+1)}, \bm z^{(t+1)})-\bm L(\bm c^{(t)}, \bm z^{(t)})\right|}{\bm L(\bm c^{(t)}, \bm z^{(t)})} < \epsilon $}
	   \STATE break
	   \ENDIF
	   \STATE $t \leftarrow t+1$
       \ENDWHILE
       \ENSURE return $\hat{\bm c}=\bm c^{(t)}$, $\hat{\bm z}=\bm z^{(t)}.$
	\end{algorithmic}
\label{algo_TSDC}
\end{algorithm}

\section{Consistency Results}\label{section:theory}

In this section, we demonstrate the identifiability of the community structure under the TNPM model and establish the consistency of the DOM algorithm. This includes the consistency of estimating the connection probability matrix and the consistency of community detection. For generality, throughout this section, we assume that each entry $A_{ij}$ of the adjacency matrix $A$ follows a sub-Gaussian distribution with variance proxy $\sigma_{ij}^{2}$ when $\bm c$ and $\bm z$ are given. Specifically, given $\bm c$ and $\bm z$, it has
\begin{equation}
A_{ij} - \mathbb{E}(A_{i,j} | \bm c, \bm z) \sim subG(\sigma_{ij}^{2}), \quad \text{for any $i \in [n]$, $j \in [m]$} ,
\label{adj_subGaussian}
\end{equation}
where $\sigma_{ij}^{2}$ satisfies that
\begin{equation}
\mathbb{E}[\exp (t \{A_{ij} - \mathbb{E}(A_{i,j} | \bm c, \bm z) \})] \leq \exp \left(\frac{\sigma_{ij}^2 t^2}{2}\right), \quad \forall t \in \mathbb{R}.\nonumber
\end{equation}
Note that $\operatorname{subG}\left(\sigma^2\right)$ denotes a class of distributions rather than a distribution. Therefore, the notation is slightly abused when writing some random variable $X \sim \operatorname{subG}\left(\sigma^2\right)$. All the proofs and
technical details are presented in the Appendix A and B.

\subsection{Identifiability of Community Structure}

We first demonstrate the identifiability of community structures under the TNPM model, assuming numbers of communities, $K$ and $L$, are known. According to the model setting, the parameters $\Lambda$ and $\widetilde{\Lambda}$ have the following structure
\begin{eqnarray}
\Lambda=\left(
\begin{array}{cccc}
\Lambda_{11} & \Lambda_{12} & \ldots  & \Lambda_{1L}\\
\Lambda_{21} & \Lambda_{22} & \ldots  & \Lambda_{2L}\\
\vdots & \vdots & &\vdots \\
\Lambda_{K1} & \Lambda_{K2} & \ldots  & \Lambda_{KL}\\
\end{array}
\right), \;
\widetilde{\Lambda}=
\left(
\begin{array}{cccc}
\widetilde{\Lambda}_{11} & \widetilde{\Lambda}_{12} & \ldots  & \widetilde{\Lambda}_{1K}\\
\widetilde{\Lambda}_{21} & \widetilde{\Lambda}_{22} & \ldots  & \widetilde{\Lambda}_{2K}\\
\vdots & \vdots & &\vdots \\
\widetilde{\Lambda}_{L1} & \widetilde{\Lambda}_{L2} & \ldots  & \widetilde{\Lambda}_{LK}\\
\end{array}
\right). \nonumber
\end{eqnarray}
To analyze the identifiability of the community structure under TNPM, we make the following assumptions:

Assumption A1: All of the elements in $\Lambda$ and $\widetilde{\Lambda}$ are positive.

Assumption A2: The points in the same community are in $general\; position$, which implies that any subset of $L$ rows of the matrix $\Lambda_{k}=\left[\Lambda_{k1}, \Lambda_{k2}, \ldots, \Lambda_{kL}\right]$ are linearly independent for any $k \in [K]$, and any subset of $K$ rows of the matrix $\widetilde{\Lambda}_{l}=\left[\widetilde{\Lambda}_{l1}, \widetilde{\Lambda}_{l2}, \ldots, \widetilde{\Lambda}_{lK}\right]$ are linearly independent for any $l \in [L]$.

Assumption A3: $n\geq K^{2}L$ or $m\geq L^{2}K$.

\noindent
$\mathbf{Remark \ 1:}$
Assumptions A1, and A2 impose conditions and restrictions on the parameters $\Lambda$ and $\widetilde{\Lambda}$, and have also been employed in prior works such as \cite{sengupta2018block} and \cite{noroozi2021estimation}. Assumption A3 introduces the lower bound of the network scale ($n$, $m$) according to the number of communities, $K$ and $L$. Given that real-world networks often consist of a large number of nodes, while the number of communities is typically small, this assumption is naturally satisfied. 
\noindent
\begin{theorem} 
  \label{identifiability}
 Under the TNPM, assuming that Assumptions A1$\sim$A3 hold, we consider the following optimization problem
 \begin{eqnarray}
 \left(\hat{\c}, \hat{\z} \right)=\arg\min_{\c \in [K]^{n}, \z \in [L]^{m}}Loss(\c, \z),
 \nonumber
\end{eqnarray}
  with
 \begin{eqnarray}
 Loss(\c, \z)&=&\sum\limits_{k=1}^{K}\sum\limits_{l=1}^{L}\left\|P^{(k,l)}((\bm c, \bm z)-\Pi_{1}\left\{P^{(k,l)}(\bm c, \bm z)\right\}\right\|_{F}^{2},\nonumber\\
 &=&\min_{\substack{\lambda_{kl}, k\in [K], l \in [L] \\\mu \in R^{K \times n}, \tilde{\mu}\in R^{m \times L}}}\sum\limits_{k=1}^{K}\sum\limits_{l=1}^{L}
 \left\|P^{(k,l)}(\bm c, \bm z) - \lambda_{kl}\tilde{\mu}_{[\c=k], l}\mu_{k, [\z=l]}\right\|_{2}^{2},\nonumber
 \end{eqnarray}
 where $\bm \c$,$\bm \z$ represent the clustering vectors. Then, we have $\hat{\c} \equiv\c^{\ast}\;\mathrm{and}\; \hat{\z} \equiv\z^{\ast}\nonumber$, where $\bm c^{\ast}$ and $\bm z^{\ast}$ are the ground truth community structures, and $\equiv$ indicates that the two community label assignments on both sides coincide up to a permutation $\bm{\pi}$ on $\{1, 2, \ldots, K\}$ or $\{1, 2, \ldots, L\}$.
 \end{theorem}

\noindent
$\mathbf{Remark \ 2:}$
Theorem \ref{identifiability} provides the conditions for identifiability and demonstrates that under these assumptions, the ground truth community structures $\bm {c}^{\ast}$ and $\bm {z}^{\ast}$ can be uniquely determined by the mean structure of the adjacency matrix $P=\mathbb{E}(A)$.

\subsection{Consistency of Estimated Connection Probability Matrix}

In this subsection, we evaluate the error associated with the estimated connectivity probability matrix, which is obtained by using the DOM algorithm. The penalty term involved in the DOM algorithm is carefully chosen to exceed the random errors (See inequality (A.21) in the Appendix A). Specifically, we introduce the penalty as

\begin{equation}
\label{penalty}
Pen(n,m,K,L)=2\tilde{\sigma}_{max}^{2}
\left\{(1+1/{\alpha_{2}})F_{1}(n,m,K,L)+(1/{\alpha_{1}})F_{2}(n,m,K,L)\right\}, 
\end{equation}
where $\tilde{\sigma}_{max}^{2}$ is an absolute constant, specified in advance, and not smaller than $\sigma_{max}^{2}\triangleq \max_{i\in [n], j\in [m]}\sigma_{ij}^{2}$. Since $\sigma_{max}^{2}$ is unknown in real applications, we need to choose a sufficiently large but reasonable $\tilde{\sigma}_{max}^{2}$. $F_{1}(n,m,K,L)$ and $F_{2}(n,m,K,L)$ are defined as
\begin{eqnarray}
&&F_{1}(n,m,K,L)=C\left\{nL+mK+KL\log (2KL)+KL(n\log K+\log n+m\log L+\log m)\right\},\nonumber\\
&&F_{2}(n,m,K,L)=n\log K+\log n+m\log L+\log m.\nonumber
\end{eqnarray}
Note that the constants $\left\{\alpha_{1}, \alpha_{2}, C\right\}$ involved above are all positive values that can be calculated, given in the proof of the following Theorem \ref{consistency_of_connection_probability}.
 \begin{theorem}
 \label{consistency_of_connection_probability}
 Under the TNPM with $\sigma_{max}^{2}\triangleq \max_{i\in [n], j\in [m]}\sigma_{ij}^{2}$, let $\left(\widehat{K}, \widehat{L}, \hat{\bm c}, \hat{\bm z}\right)$ defined as
 \begin{eqnarray}
 \left(\widehat{K}, \widehat{L}, \hat{\bm c}, \hat{\bm z}\right)=
 \arg\min_{(K,L,\bm c,\bm z)}\Bigg\{
 \sum\limits_{k=1}^{K}\sum\limits_{l=1}^{L}\left\|A^{(k,l)}(\bm c,\bm z)- \Pi_{1}\left\{A^{(k,l)}(\bm c,\bm z)\right\} \right\|_{F}^{2}\nonumber\\
 +Pen(n,m,K,L)\Bigg\},
 \label{definition_KLCZ}
 \end{eqnarray}
 \normalsize
 then $\left\|\widehat{P}-P_{\ast}\right\|_{F}^{2}=\sum\limits_{k=1}^{\widehat{K}}
 \sum\limits_{l=1}^{\widehat{L}}
 \left\|\Pi_{1}\left\{A^{(k,l)}(\hat{\bm c}, \hat{\bm z})\right\}-P_{\ast}^{(k,l)}(\hat{\bm c}, \hat{\bm z})\right\|_{F}^{2}
 $ satisfies the following inequalities for any $t\geq 0$ and some positive constants $H_{1}=\frac{1}{1-\alpha_{1}-4\alpha_{2}}$, $H_{2}=\frac{2C+2/\alpha_{1}+2C/\alpha_{2}}{1-\alpha_{1}-4\alpha_{2}}$:
 \begin{eqnarray}
 &&\mathbb{P}\left\{
 \frac{1}{nm}\left\|\widehat{P}-P_{\ast}\right\|_{F}^{2}
 \leq \frac{H_{1}}{nm}Pen(n,m,K_{\ast},L_{\ast})+\frac{H_{2}\sigma_{max}^{2}}{nm}t
 \right\}\geq 1-3e^{-t},\label{first_part_consistncyP}\\
 &&\frac{1}{nm}\mathbb{E}\left\|\widehat{P}-P_{\ast}\right\|_{F}^{2}
 \leq \frac{H_{1}}{nm}Pen(n,m,K_{\ast},\text{\L}_{\ast})+\frac{3H_{2}\sigma_{max}^{2}}{nm}.\label{second_part_consistncyP}
 \end{eqnarray}
 \end{theorem}

\noindent
$\mathbf{Remark \ 3:}$
Theorem \ref{consistency_of_connection_probability} guarantees the consistency of the estimated connectivity probability matrix obtained by using the DOM algorithm under the TNPM model. The estimation remains consistent when$\frac{KLlog(KL)}{n \wedge m} \rightarrow 0$. In fact, according to the equation (\ref{penalty}) and inequality (\ref{second_part_consistncyP}), $Pen(n,m,K,L)$ converges to $0$ when $\frac{KLlog(KL)}{n \wedge m} \rightarrow 0$.

\noindent
$\mathbf{Remark \ 4:}$
Theorem \ref{consistency_of_connection_probability} significantly differs from the theoretical result in \cite{noroozi2021estimation}. First, we greatly extend the applicable types of networks, including binary networks, discrete-valued networks, continuous-valued networks, and even with mixture link distributions. Second, we use a new strategy to directly upper bound the operator norm of random matrices with sub-Gaussian entries. Following \cite{noroozi2021estimation} will lead to the concentration inequality for the Lipschitz function of independent sub-Gaussian random variables, while it remains a highly challenging research topic in the academic community.

\subsection{Consistency of Community Detection}

In this subsection, we evaluate the error associated with the estimated community structure obtained by using the DOM algorithm. For the convenience of theoretical analysis, we assume that the true number of communities $K=K^{\ast}$ and $L=L^{\ast}$ is known, like \cite{noroozi2021sparse, noroozi2021estimation}.

Let $C_{\ast} \in \mathcal{M}_{n,K}$ denote the ground truth out-community matrix, $Z_{\ast} \in \mathcal{M}_{m,L}$ denote the ground truth in-community matrix. Let $C$ and $Z$ represent other out-community and in-community matrices, respectively. We define the proportion of misclassified nodes by $C$ and $Z$ as follows:
\begin{eqnarray}
&&\operatorname{Err}\left(C, C_{*}\right)=(2 n)^{-1} \min _{\mathcal{P}_{K} \in \mathbb{P}_{K}}\left\|C \mathcal{P}_{K}-C_{*}\right\|_{1}=(2 n)^{-1} \min _{\mathcal{P}_{K} \in \mathbb{P}_{K}}\left\|C \mathcal{P}_{K}-C_{*}\right\|_{F}^{2},\nonumber\\
&& \operatorname{Err}\left(Z, Z_{*}\right)=(2 m)^{-1} \min _{\mathcal{P}_{L} \in \mathbb{P}_{L}}\left\|Z \mathcal{P}_{L}-Z_{*}\right\|_{1}=(2 n)^{-1} \min _{\mathcal{P}_{L} \in \mathbb{P}_{L}}\left\|Z \mathcal{P}_{L}-Z_{*}\right\|_{F}^{2},\nonumber
\end{eqnarray}
where $\mathbb{P}_{K}$ is the set of permutation matrices $\mathcal{P}_{K}: \{1, 2, \ldots, K\} \longrightarrow \{1, 2, \dots, K\}$, and $\mathbb{P}_{L}$ is the set of permutation matrices $\mathcal{P}_{L}: \{1, 2, \ldots, L\} \longrightarrow \{1, 2, \dots, L\}$. Additionally, we define
\begin{eqnarray}
\Upsilon (C_{\ast}, Z_{\ast}, \rho_{n,m})=\left\{
(C,Z)\in \mathcal{M}_{n,K}\times \mathcal{M}_{m,L}:
\max \left\{\operatorname{Err}\left(C, C_{*}\right),  \operatorname{Err}\left(Z, Z_{*}\right)\right\}\geq \rho_{n,m}
\right\}\nonumber
\end{eqnarray}
as the set of community matrices with the proportion of misclassified nodes being at least $\rho_{n,m}\in (0,1)$. 
 \begin{theorem}
 \label{clustering_error}
 Under the TNPM with $\sigma_{max}^{2}\triangleq \max_{i\in [n], j\in [m]}\sigma_{ij}^{2}$, assuming that Assumptions A1 $\sim$ A3 hold, let $(\widehat{C}, \widehat{Z}) \equiv (\widehat{C}_{K}, \widehat{Z}_{L})$ be the community matrices corresponding to  \eqref{definition_KLCZ}. If there exist $\alpha \in (0,1/2)$ and $\rho_{n,m} \in (0,1)$ such that the following inequality holds
 \begin{eqnarray}
 &&\|P_{\ast}\|_{F}^{2}-(1+\alpha)\max_{(C,Z)\in \Upsilon (C_{\ast}, Z_{\ast}, \rho_{n,m})}
 \sum\limits_{k=1}^{K}\sum\limits_{l=1}^{L}\|P_{\ast}^{(k,l)}(C,Z)\|_{op}^{2} \\
 &\geq&\sigma_{max}^{2}\left[
H_{1}\left\{nL+mK+KL\log(2KL)+KL(m+n)\right\}+H_{2}KL(n\log K+m\log L)\right],\nonumber
 \label{outlier_Set}
 \end{eqnarray}
 where $\alpha$, $H_{1}$ and $H_{2}$ are absolute positive constants with their definitions provided the proof of this theorem. Then, with probability at least $1-2e^{-(n+m)}$, the proportion of nodes misclassified by $(\widehat{C}, \widehat{Z})$ is at most $\rho_{n,m}$, i.e.,
 \begin{eqnarray}
 \max \left\{\operatorname{Err}\left(C, C_{*}\right),  \operatorname{Err}\left(Z, Z_{*}\right)\right\}\leq \rho_{n,m}.
 \label{upper_bound}
 \end{eqnarray}
 \end{theorem}

\noindent
$\mathbf{Remark \ 5:}$ The condition (\ref{outlier_Set}) means that if the community matrices $(C, Z)$ fall within the set where the proportion of misclassified nodes is at least $\rho_{n,m}$, there will exist a lower bound on the sum of the differences between the Frobenius and operator norms of the blocks  $\left\{P_{\ast}^{(k,l)}(C,Z)\right\}$. In fact, if the clustering is incorrect, the ranks of the blocks would increase which would result in a discrepancy between their operator and Frobenius norms. 

\noindent
$\mathbf{Remark \ 6:}$ 
Theorem \ref{clustering_error} provides an upper bound on the misclassification rate, going beyond the conventional statement that it tends to zero as the network size increases, as it is routinely done in papers that rely on modularity maximization for clustering assignments (see, e.g. \cite{bickel2009nonparametric, zhao2012consistency,  sengupta2018block}). Similar conclusion for undirected networks is obtained in \cite{noroozi2021estimation}. To the best of our knowledge, this is the cutting-edge result available so far.

\section{Simulation Studies}\label{section:simulation}

In this section, we evaluate the performance of our proposed methods using synthetic networks, concentrating on two main aspects: the accuracy of community detection and computational efficiency. The code is publicly available at Github
(\href{https://github.com/Wangya1996/Two-way-Node-Popularity-Model}{https://github.com/Wangya1996/Two-way-Node-Popularity-Model}).

\subsection{Accuracy of Community Detection}

The adjacency matrix $A$ is generated element-wisely using Normal, Bernoulli, and Poisson distributions, guided by a probability matrix $P$ defined by the TNPM model. Additionally, a mixture of Normal and Bernoulli distributions is examined to validate our theoretical findings in sub-Gaussian contexts. We only show the Normal and Normal-Bernoulli mixture generation cases here, and the rest are in the Appendix C.. 

The pair of community number $(K,L)$ are set to be (3,4). We first assume that $K$ and $L$ are known, and will later consider the estimation of $K$ and $L$. The elements of block matrices $\Lambda$ and $\widetilde{\Lambda}$ are drawn independently from $U[0,1]$. The ground truth out-community assignments $\bm c \in [K]^n$ and in-community assignments $\bm z \in [L]^m$ are generated from a multinomial distribution, such that $P(\bm c_i=k)=1/K$ and $P(\bm z_j=l)=1/L$, where $k \in [K]$ and $l \in [L]$. Furthermore, for evaluating our methods in sparse data situations, we introduce a sparsity parameter $\eta$ as the proportion of nonzero entries in matrix $\Lambda$ and $\widetilde{\Lambda}$. To induce sparsity, we set the $\lfloor nL\eta \rfloor$ ($\lfloor mK\eta \rfloor$) smallest non-diagonal entries of $\Lambda$ ($\widetilde{\Lambda}$) to zero. 

We evaluate the performance of our proposed methods in comparison to current state-of-the-art approaches:
OMPSC: A sparse subspace clustering method introduced by \cite{noroozi2021estimation};
COSSC and INSC: Both methods employ spectral clustering techniques, with COSSC leveraging a cosine similarity-based matrix and INSC utilizing an inner product similarity matrix;
SVDK: This approach implements the K-means algorithm on the singular matrices derived from the network's adjacency matrix.

Note that OMPSC, COSSC, and INSC are fundamentally developed for symmetric networks. To facilitate their application in analyzing directed or bipartite networks, we separately apply these methods to both the network's adjacency matrix and its transpose. All the simulation results are based on 100 independent replications.

For community detection, we compare the performance of the DOM and TSDC algorithms against other techniques using three metrics: the clustering error in \cite{wang2010consistent} and \cite{zhang2021directed}, the normalized mutual information (NMI) in \cite{lancichinetti2009detecting} and \cite{zhou2020optimal}, and the proportion of misclustered nodes in \cite{noroozi2021sparse,noroozi2021estimation}. Details on these metrics are accessible in Appendix C.1. This section primarily highlights the results using the NMI metric, while additional metrics results are provided in Appendix C.1.

We explore three distinct scenarios and the simulation settings are outlined as follows:
\begin{itemize}
    
    \item \textit{Normal case}:
        The node counts $(n, m)$ are set as$(600, 600)$. The adjacency matrix $A=(A_{ij})$ is generated with entries $A_{ij} \sim \mathcal{N}(P_{ij}, \sigma^2)$, and $\sigma$ is varied from 0 to 0.6 with the increment of 0.1.
    \item \textit{Normal-Bernoulli Mixture case}: 
        The values of $n$ and $m$ range from $360$ to $1320$, increasing in increments of $240$. The lower half of the adjacency matrix $A$ is filled with Bernoulli variables $A_{ij} \sim \text{Ber}(P_{ij})$ for $i \in [n]$ and $j = 1, \ldots, i-1$. When $j > i$, entries follow a normal distribution $A_{ij} \sim \mathcal{N}(P_{ij}, \sigma^2)$, with $\sigma$ fixed at 0.1.
    \item \textit{Sparse case}: 
        The values of $n$ and $m$ range from $200$ to $1000$, increasing in increments of $100$, while the sparsity parameter $\eta$ is chosen from $\{0.3,0.5,0.7\}$. The adjacency matrix $A$ is generated with  Bernoulli variables $A_{ij} \sim \text{Ber}(P_{ij})$.
\end{itemize}

\begin{figure}[!ht]
\centering
\subfigcapskip=-6pt
\subfigtopskip=2pt
\subfigbottomskip=0pt
\subfigure[out-community]{{}
\includegraphics[width=7cm,height=4.8cm]{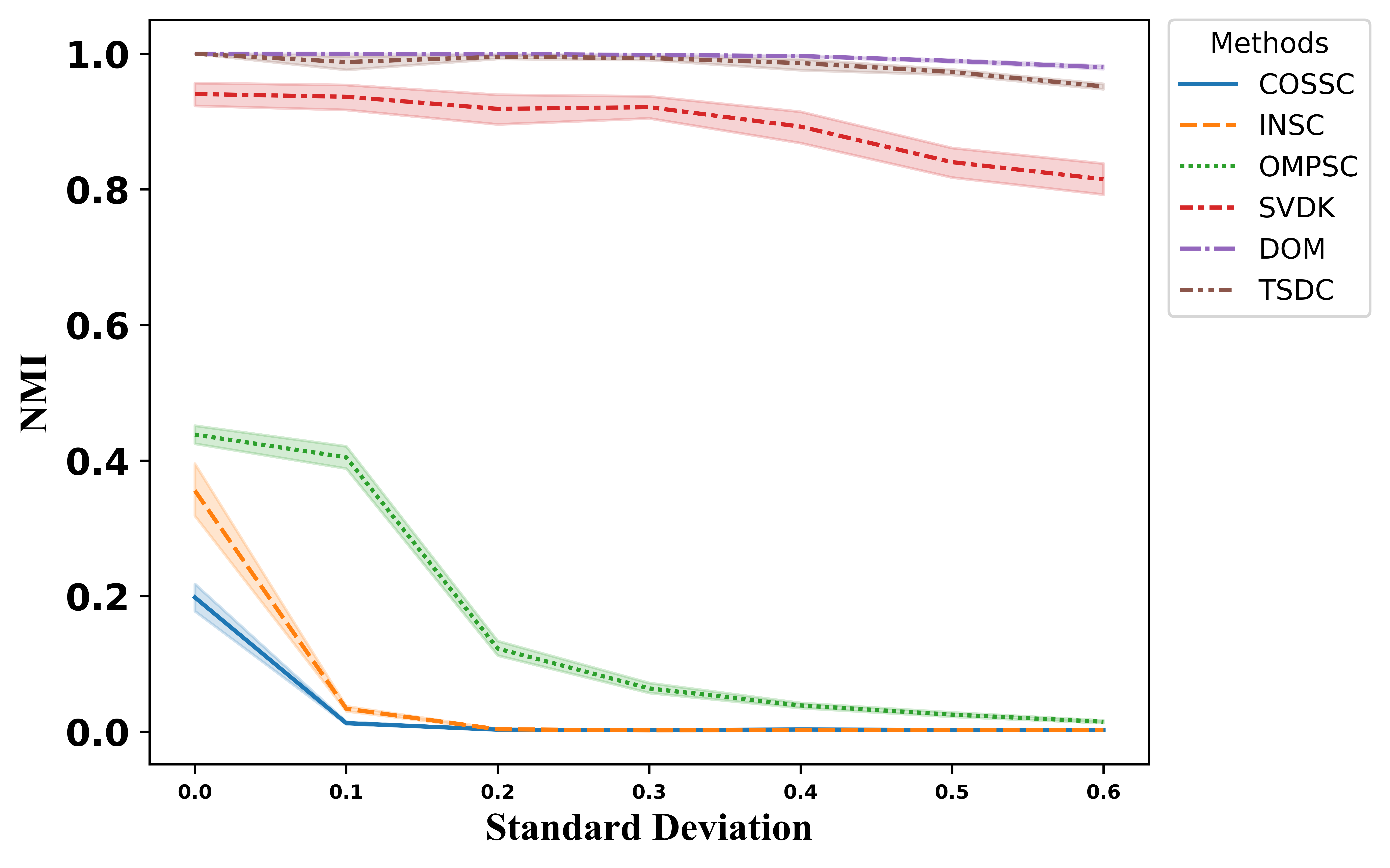}}
\hspace{0.01\linewidth}
\subfigure[in-community]{{}
\includegraphics[width=7cm,height=4.8cm]{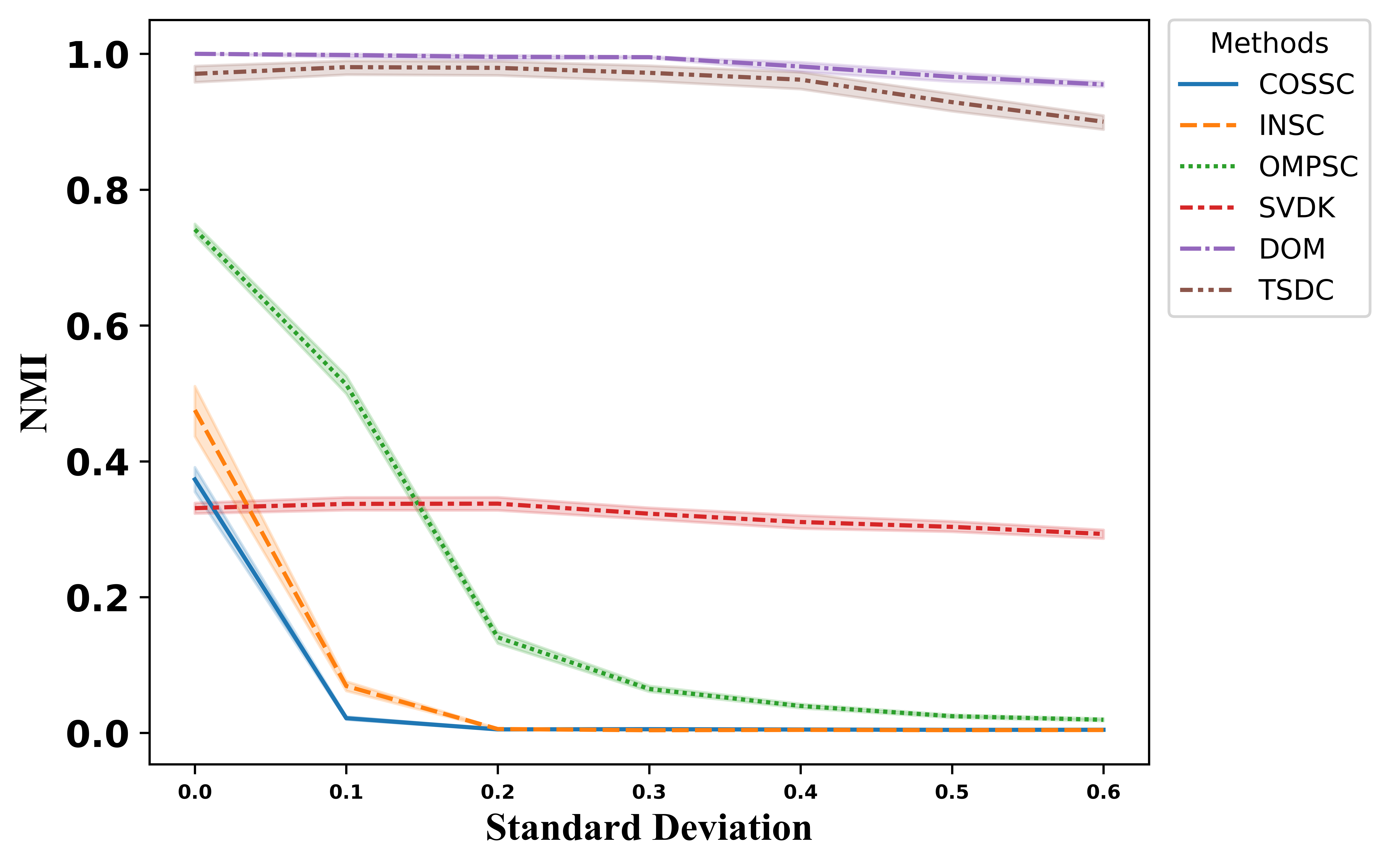}}\vspace{0.02\linewidth}
\caption{The NMI for Normal data generation case with $(n,m)=(600,600)$. The left panel depicts out-community clustering, while the right panel shows in-community clustering.}
\label{Normal_fig}
\end{figure}

\begin{figure}[!ht]
\centering
\subfigcapskip=-6pt
\subfigtopskip=2pt
\subfigbottomskip=0pt
\subfigure[out-community]{{}
\includegraphics[trim=0 -1mm 0 0, width=7cm,height=4.8cm]{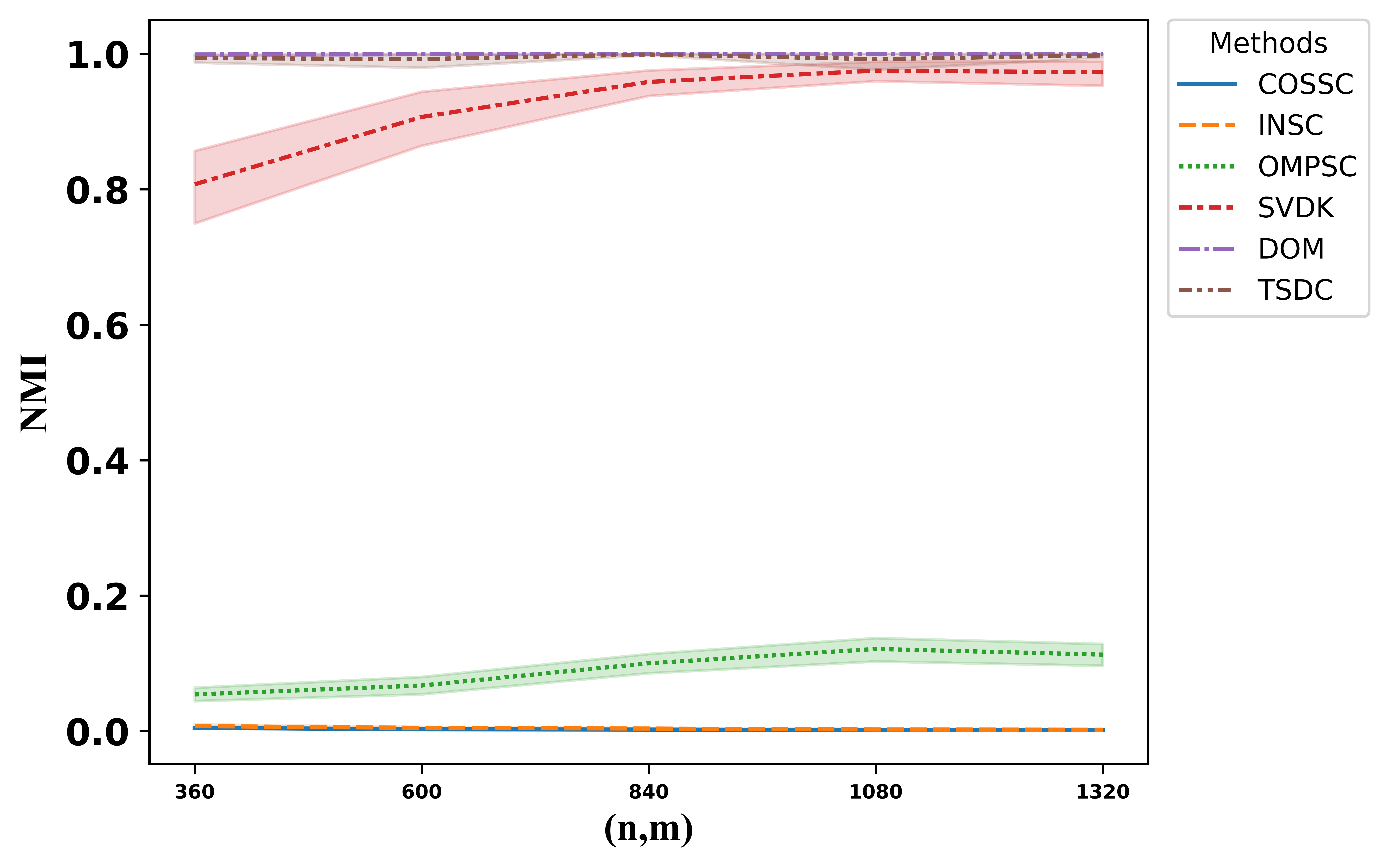}}
\hspace{0.01\linewidth}
\subfigure[in-community]{{}
\includegraphics[trim=0 -1mm 0 0,width=7cm,height=4.8cm]{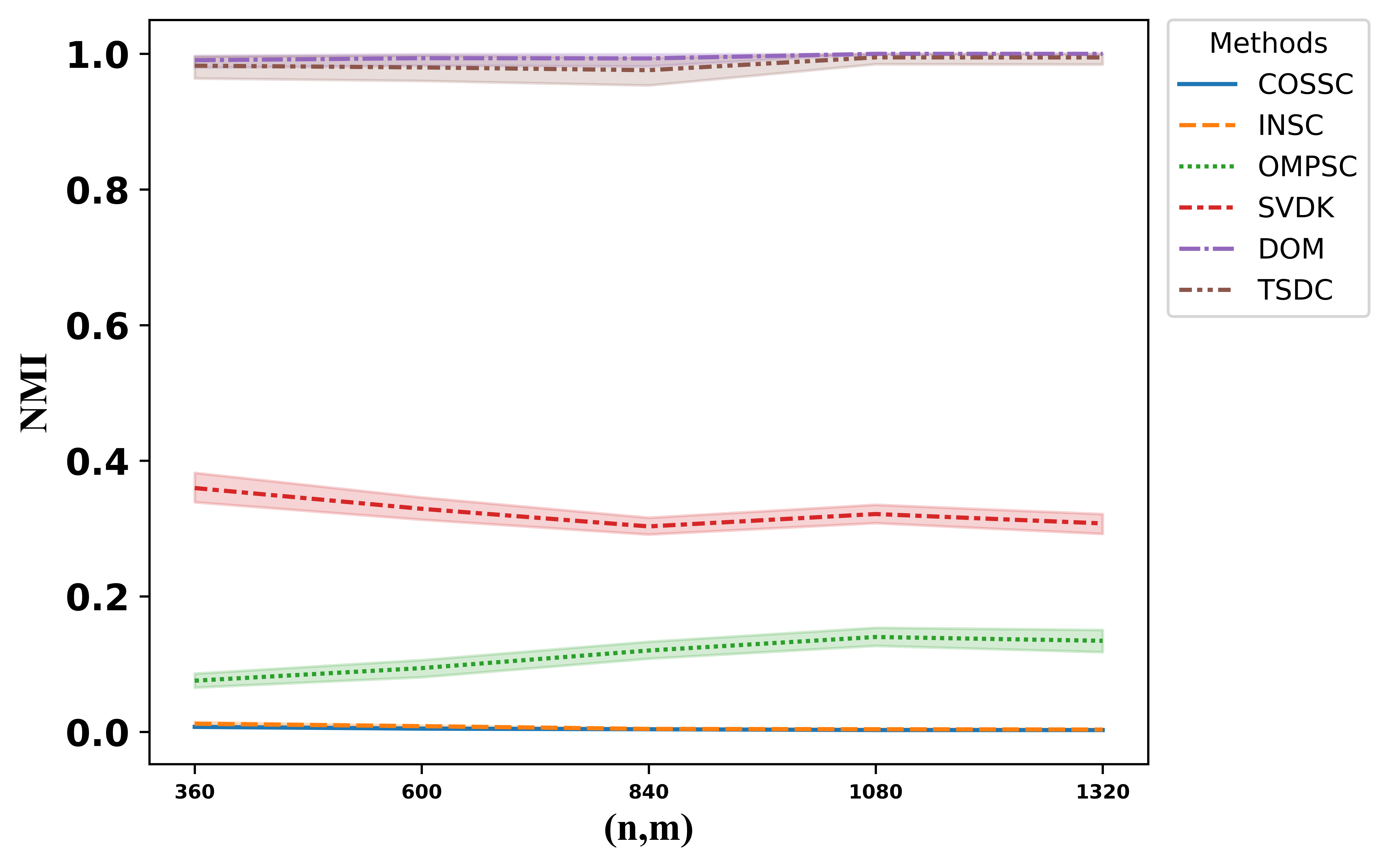}}\vspace{0.02\linewidth}
\caption{The NMI for Normal-Bernoulli mixture data generation case. The left panel depicts out-community clustering, while the right panel shows in-community clustering.}
\label{Mixture_fig}
\end{figure}

\begin{figure}[!ht]
\centering
\includegraphics[trim=30mm 10mm 0 0,width=12cm, height=4cm]{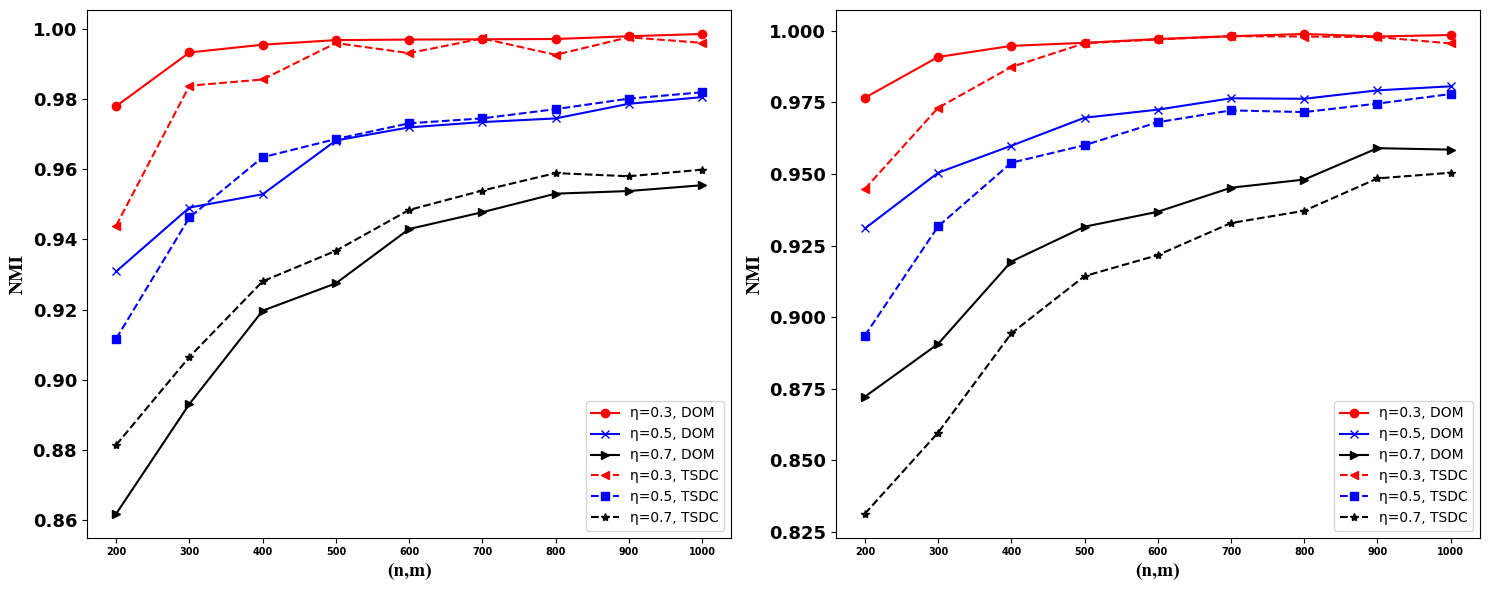}
\caption{The NMI for sparsity data generation case. The left panel depicts out-community clustering, while the right panel shows in-community clustering.}
\label{Spar_NMI}
\end{figure}

Figures \ref{Normal_fig} and \ref{Mixture_fig} clearly demonstrate that the DOM and TSDC algorithms exceed the performance of other methods. Appendix C.1 corroborates these findings, presenting consistent outcomes via the metrics of Misclassification Score (MIS) and Clustering Error. 

From Figure \ref{Spar_NMI}, an increase in sparsity ($\eta$ increases) correlates with a decrease in NMI. Additionally, the DOM method outperforms the TSDC method, especially in scenarios involving smaller networks and lower levels of sparsity. However, as the network size grows, the distinction between the DOM and TSDC results narrows.

\subsection{Computational Efficiency}

In this section, We assess the computational efficiency of each method by measuring the running time (in seconds). All methods are implemented in Matlab and run on a single processor of an Intel(R) Core(TM) i9-12900K CPU 3.20 GHz PC.

We set the values of $n$ and $m$ range from 360 to 1320, increasing in increments of 240 and generate the adjacency matrix $A=(A_{ij})$ where $A_{ij} \sim \mathcal{N}(P_{ij}, \sigma^2)$ with $\sigma$ is chosen from $\{0.1,0.5\}$. The computational efficiency of each algorithm, based on the average runtime from 100 simulations, is summarized in Table \ref{tab1}. The results reveal that the TSDC algorithm significantly outperforms the DOM algorithm in terms of processing speed,and a finding echoed in Appendix C.1 for matrices generated from Bernoulli and Poisson distributions. As noise levels rise, so does the computation time. This suggests the DOM algorithm as a viable option for smaller networks or non-urgent processing, whereas TSDC stands out for larger networks or scenarios requiring quick processing.

\begin{table}[H]
\renewcommand{\arraystretch}{1.3}
\centering
\begin{tabular}{c | c | c c c c c c}
\hline
\hline
$\sigma$ & $(n,m)$& DOM & TSDC & OMPSC & COSSC & INSC & SVDK\\
\hline
\multirow{4}{*}{0.1}& (360,360) &45.44 & 0.07 & 0.90& 0.11& 0.19&0.03 \\
\cline{2-8}
& (600,600)&173.12 &0.13  & 2.56& 0.25& 0.64& 0.06\\
\cline{2-8}
& (840,840)& 421.12&0.21  & 5.84& 0.47& 1.45&0.10\\
\cline{2-8}
&(1080,1080)&892.90 & 0.27 &11.08 &0.78 & 2.95&0.15\\
\cline{2-8}
&(1320,1320)& 1482.88& 0.57 & 14.34& 0.97& 5.41& 0.19\\
\hline
\hline
& (360,360) &66.45 & 0.27 & 0.93& 0.11& 0.19&0.03\\
\cline{2-8}
& (600,600)& 249.91& 0.26& 2.67& 0.26& 0.64& 0.06\\
\cline{2-8}
0.5& (840,840)&592.11 & 0.37 & 5.96& 0.49& 1.46&0.10\\
\cline{2-8}
&(1080,1080)&1219.89 &0.49  & 11.30& 0.80&2.94 &0.15\\
\cline{2-8}
&(1320,1320)& 1657.89& 0.59 & 14.87&0.98 & 5.38& 0.19\\
\hline
\hline
\end{tabular}
\caption{Running time (in seconds) for scenarios with $n$ and $m$ varying from 360 to 1320 in increments of 240, under the Normal data generation case for $\sigma$=0.1,0.5.}
\label{tab1}
\end{table}

\subsection{Unknown number of clusters}

In our previous simulations, the true number of clusters $K$ and $L$ are given. If not, they can be estimated as solving the optimization problem (\ref{eqn:(13)}), which can be equivalently rewritten as 
\begin{equation}
\label{num_clusters}
\left( \widehat{K}, \widehat{L}\right)=\arg \min_{ K, L}\left\{\left \|A-\widehat{P}\right \|_F^2+Pen(n, m, K, L)\right\}.
\end{equation}
We study the selection of unknown $K$ ($L$) using an empirical version of this penalty
\begin{equation}
Pen(n,m,K,L)=\rho(A)\{nL\sqrt{\ln{n}{\ln{L}}^3}+mK\sqrt{\ln{m}{\ln{K}}^3}\}
\end{equation}
Table 3, 4 in Appendix C.3 present the relative frequencies of the estimator $(\hat{K},\hat{L})$ of $(K,L)$ chosen from $\{(3,3),(4,3),(4,4)\}$, $n=360$ and $\sigma$=0.1,0.3,0.5. It corroborates that, in the majority of cases, $(\hat{K},\hat{L})=(K, L)$, demonstrating that the estimation method reliably identifies the number of clusters with high probability.

\section{Real Data Applications}\label{section:real}

In this section, we apply the DOM and TSDC algorithms to analyze two real datasets: the Worldwide Food Trading Networks data set and the MovieLens 100K data set.

\subsection{Worldwide Food Trading Networks}

The Worldwide Food Trading Networks is collected by \cite{de2015structural} and is available at \url{http://www.fao.org}. Our analysis focuses on trading data in 2010, specifically on two product categories: cereals and cigarettes. We exclude countries with negligible trading volume (the first quantile), remaining 93 countries. Following the logarithmic transformation, we derive two directed 93 $\times$ 93 networks.

The DOM and TSDC algorithms are implemented on two networks, with parameters K and L set as 2 and 3, respectively. The results obtained from the DOM algorithm are presented here, and TSDC's are in Appendix D.1.

\begin{figure}[!t]
\centering
\includegraphics[trim=0 15mm 0 0, width=16cm,height=5.8cm]{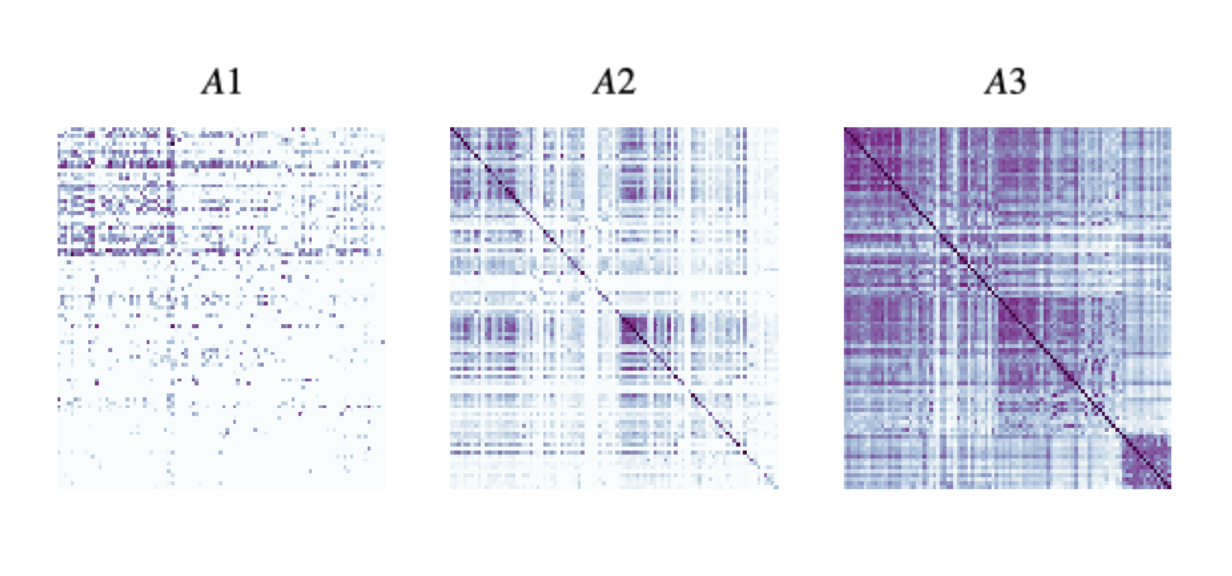}
\caption{$A1$ displays the original cereal trading network, while $A2$ and $A3$ represent the block cosine similarity matrix for the rows and columns, respectively,  with nodes ordering based on detected clustering labels.}
\label{ROA Cereals Similarity Matrix}
\end{figure}

\begin{figure}[!ht]
\centering
\includegraphics[trim=0 15mm 0 0, width=16cm,height=6cm]{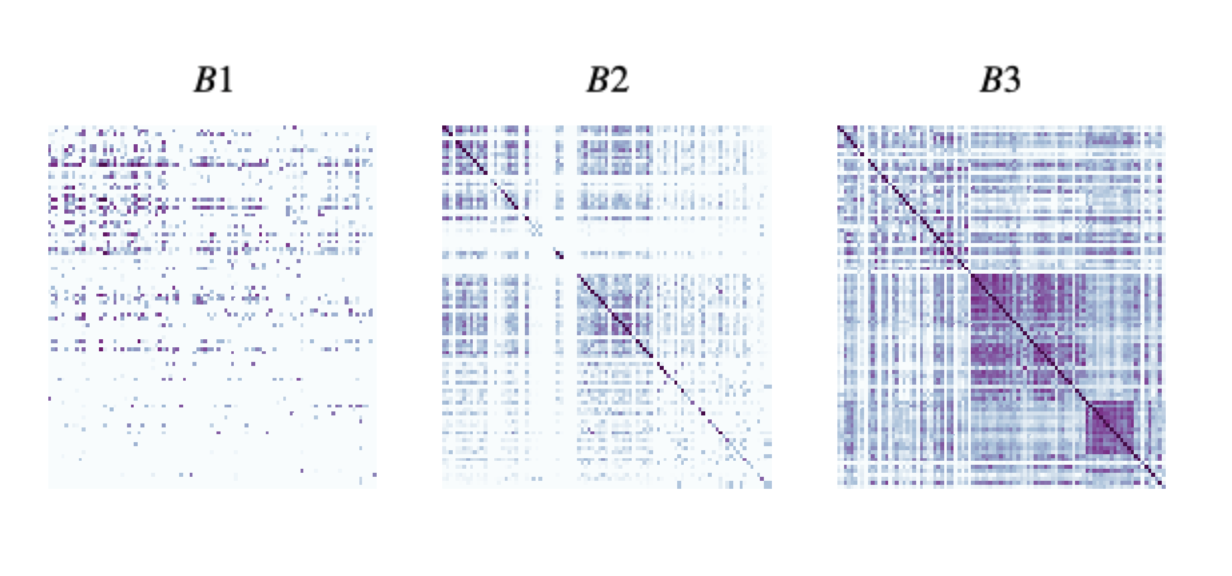}
\caption{$B1$ displays the original trading cigarette network, with Figures B2 and B3 showcasing the reordered row and column block cosine similarity matrix. }
\label{ROA Cigarettes Similarity Matrix}
\end{figure}

Figures \ref{ROA Cereals Similarity Matrix} and \ref{ROA Cigarettes Similarity Matrix} showcase the original network's heatmap alongside the reordered block cosine similarity matrix obtained by the DOM algorithm. The block cosine similarity matrix reveals distinct block patterns for exporting and importing countries, effectively demonstrating the TNPM's capacity to delineate network communities- export countries split into two groups and import countries into three, respectively.

\begin{figure}[!t]
\subfigcapskip=-8pt
\subfigtopskip=2pt
\subfigbottomskip=0pt
\subfigure[Cigarattes (Export)]{
          \includegraphics[width=7cm]
          {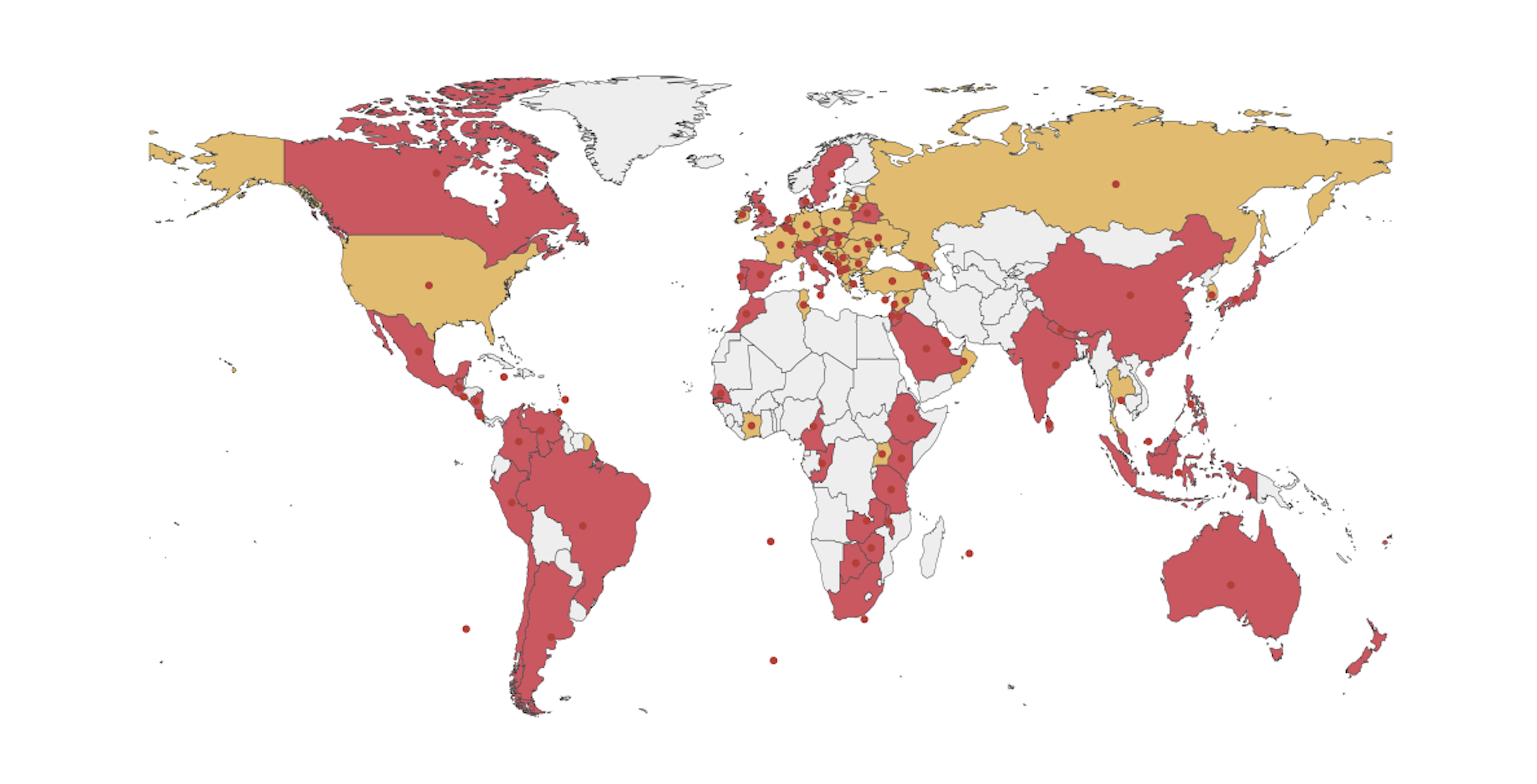}
}
\hspace{-8mm}
\subfigure[Cigarattes (Import)]{
          \includegraphics[width=7cm]
          {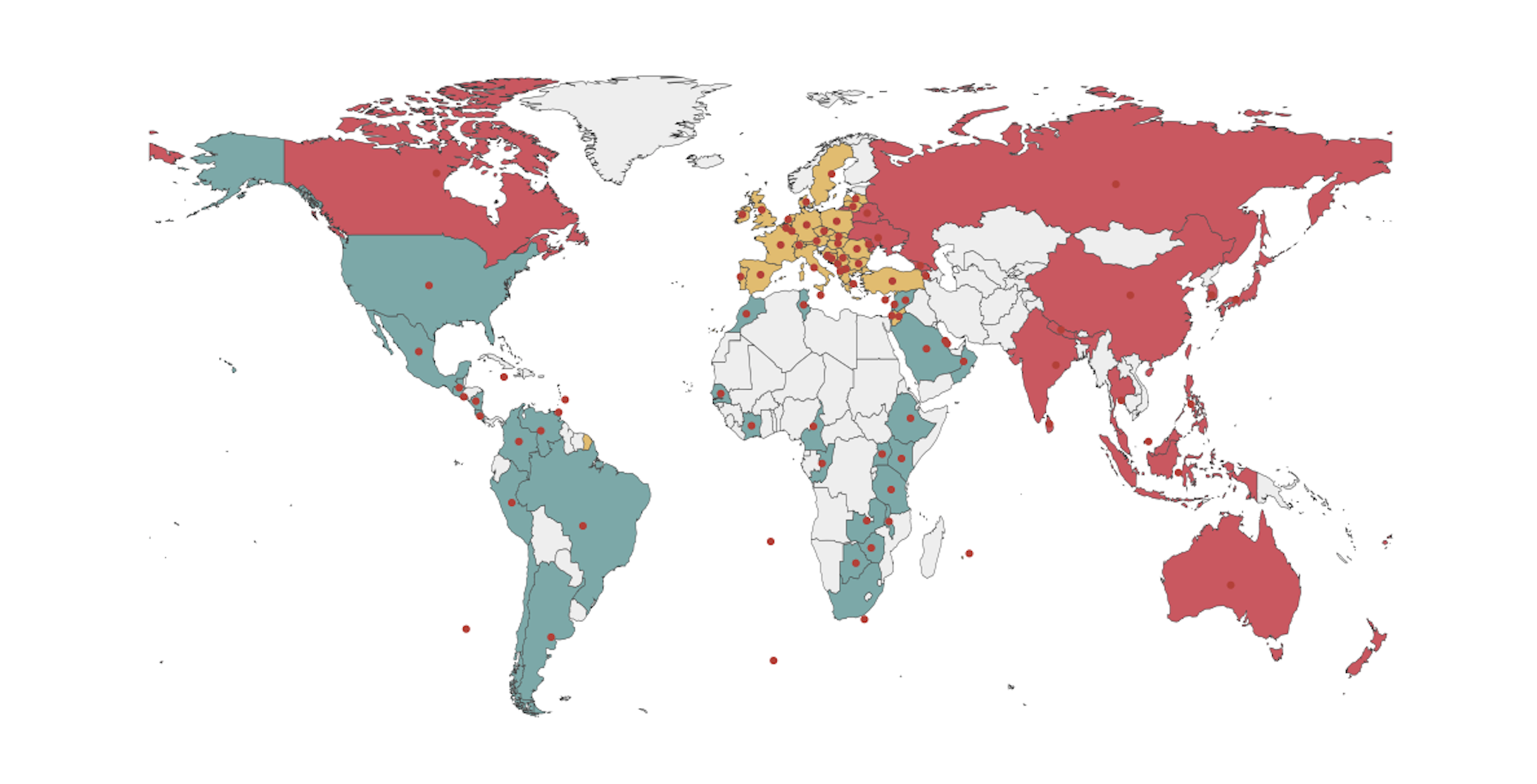}
}

\subfigure[Cereals (Export)]{
           \includegraphics[width=7cm]
           {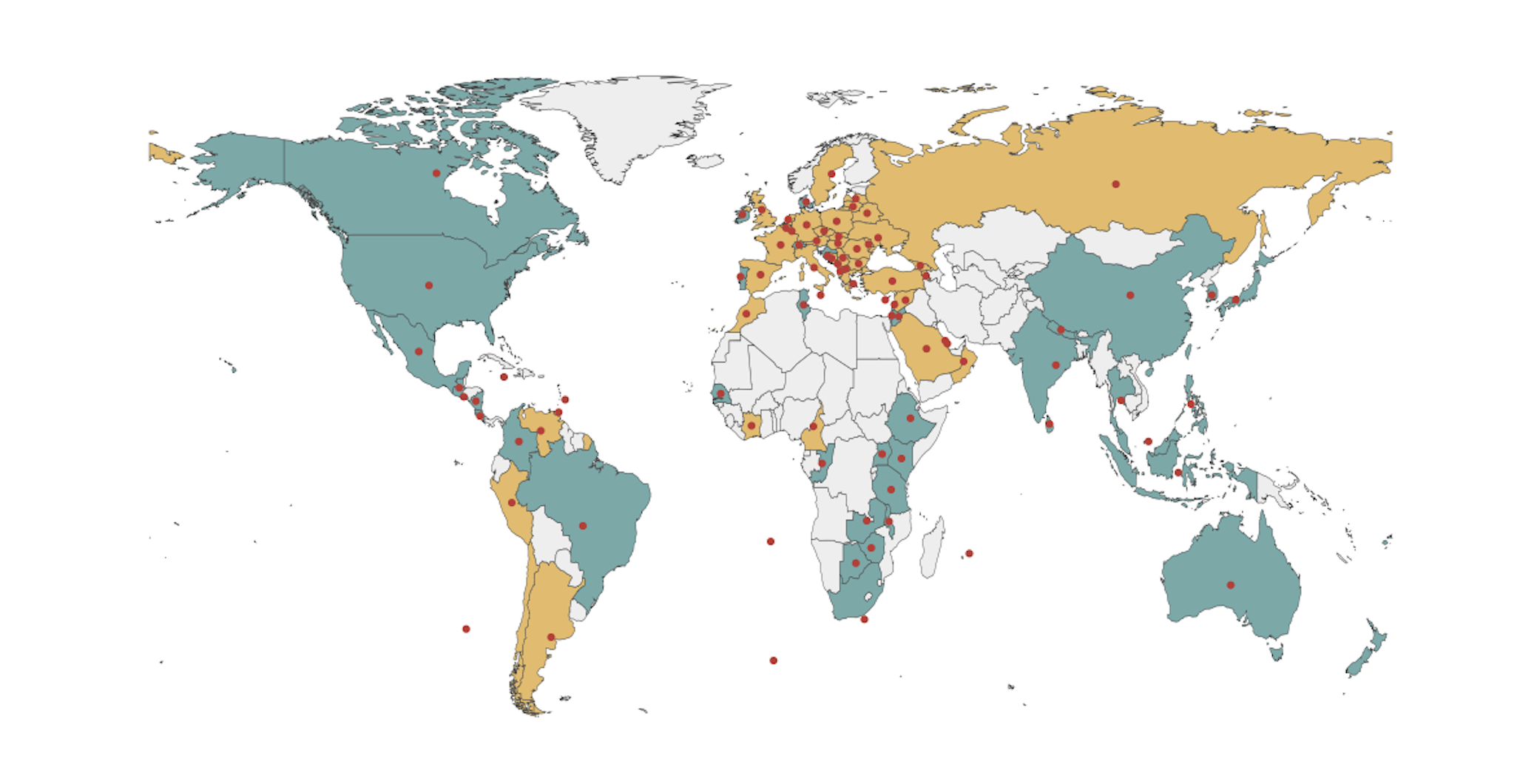}
}
\hspace{-8mm}
\subfigure[Cereals (Import)]{
           \includegraphics[width=7cm]
           {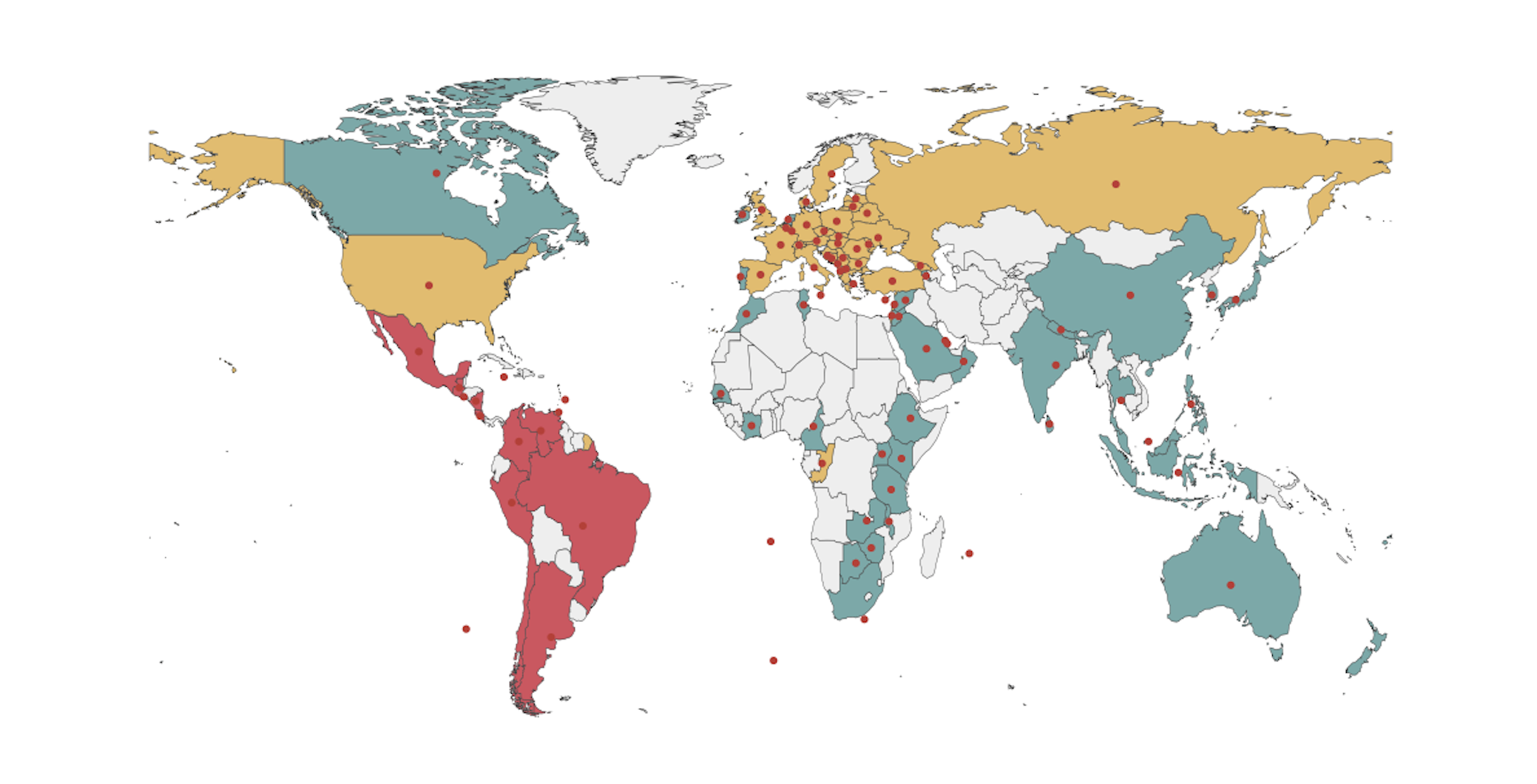}
}
\hfill
\caption{
The World Maps display: (a) cigarette export clustering, (b) cigarette import clustering, (c) cereal export clustering, and (d) cereal import clustering.}
\label{fig: ROA World Trade Map}
\end{figure}

Figure \ref{fig: ROA World Trade Map} displays country clusters with identical colors marking the same cluster and grey areas indicating excluded countries. Figure \ref{fig: ROA World Trade Map} (b) showcases country clusters in cigarette imports, revealing significant regional trade patterns. Specifically, European nations emerge as a major cluster of tobacco importers, while another cluster includes China, India, Indonesia, and other Southeast Asian and Oceanian countries, mainly sourcing tobacco from Brazil, the United States, Canada, and Argentina. A distinct cluster comprises the United States, and certain South American, and African countries, highlighting the strategic advantages of regional trading, such as lower transportation costs and quicker delivery times.

\subsection{MovieLens 100K Dataset}\label{MovieLens_realdata}

The MovieLens 100K data set, documented by \cite{harper2015movielens}, is collected by the GroupLens Research of the MovieLens website (\url{movielens.umn.edu}), and is accessible at \url{https://grouplens.org/datasets/movielens/100k}. This data set contains 10,000 ratings from 943 users across 1682 movies, leading to the construction of a 943 $\times$ 1682 rating matrix $A$, where each element $A_{i,j}$ denotes the rating from 1 to 5 given by user $i$ to movie $j$. The movies are categorized into 19 genres, including ``Adventure," ``Action," and ``Animation," among others, with 833 movies categorized in a single genre and the rest in multiple genres.

We employ the proposed algorithm to bicluster the MovieLens 100K data set. Following the parameters set forth in \cite{flynn2020profile} and \cite{zhao2024variational}, we establish $K=3$ for user clusters and $L=4$ for movie clusters. We present the TSDC's results here and leave the DOM's in Appendix D.2.

\begin{figure}[!t]
\centering
\includegraphics[width=15cm,height=5.2cm]{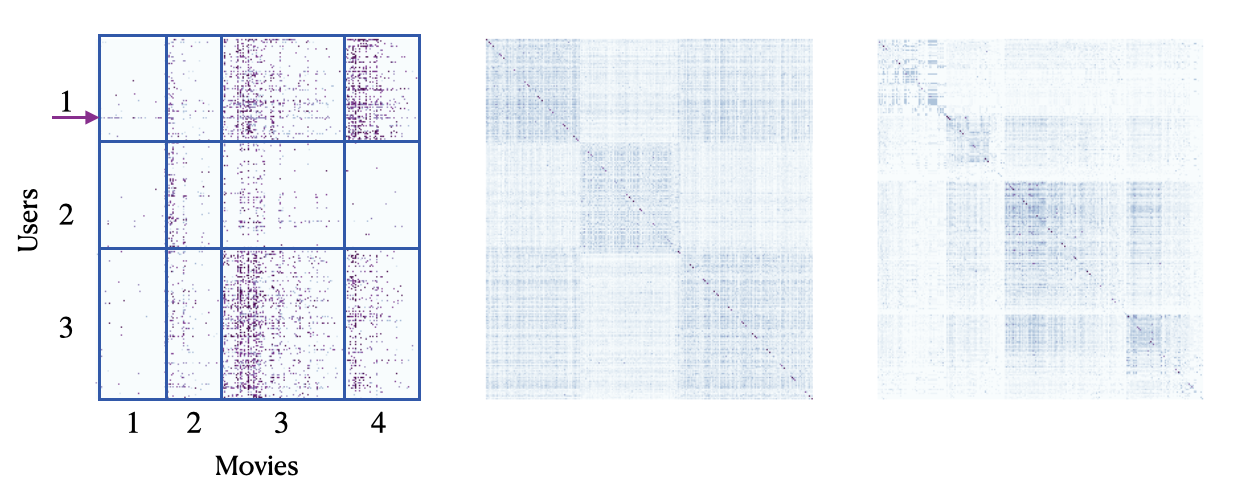}
\caption{The heatmap of the MovieLens matrix $A$ (left) and the rearranged block cosine similarity matrix for rows (middle) and columns (right).}
\label{TSDC Weighted MovieLens}
\end{figure}

Figure \ref{TSDC Weighted MovieLens} displays the data matrix heatmap and block cosine similarity matrix, with nodes organized by TSDC algorithm-detected community assignments. The left panel reveals that nodes within the same user cluster exhibit diverse patterns of node popularity across different movie clusters. For instance, in user cluster 1, one individual shows a notable preference for cluster 1 movies, whereas the rest of user cluster 1 members provide significantly fewer ratings for these films. Furthermore, our results reveal distinct consumer behavior patterns. Significantly, users in cluster 3 predominantly review movies within cluster 3, while movies in cluster 4 are primarily reviewed by users in cluster 1. The middle and right panels clearly illustrate the distinct block structures and emphasize the suitability of the TNPM for modeling the data set.

We investigate the association between estimated movie clusters and actual movie categories provided in the MovieLens data set. The direct comparison faced challenges due to the vast array of movie categories and their overlaps. To address this, we filtered the data set to 833 films, each in a unique category. We perform the chi-squared test of independence on the contingency tables and obtain p-values of 4.39 $\times$ $10^{-8}$ and 3.003 $\times$ $10^{-11}$ for the clusters estimated by the DOM and TSDC algorithms, respectively. These p-values are smaller than the reported testing p-values in \cite{flynn2020profile} (0.0415) and \cite{zhao2024variational} (2.656 $\times$ $10^{-7}$), indicating a stronger association between the algorithm-predicted clusters and the true movie categories.

\clearpage
\vskip 0.2in
\bibliography{main}

\begin{thebibliography}{24}
\providecommand{\natexlab}[1]{#1}
\providecommand{\url}[1]{\texttt{#1}}
\expandafter\ifx\csname urlstyle\endcsname\relax
  \providecommand{\doi}[1]{doi: #1}\else
  \providecommand{\doi}{doi: \begingroup \urlstyle{rm}\Url}\fi

\bibitem[Amini et~al.(2013)Amini, Chen, Bickel, and Levina]{amini2013pseudo}
Arash~A Amini, Aiyou Chen, Peter~J Bickel, and Elizaveta Levina.
\newblock Pseudo-likelihood methods for community detection in large sparse
  networks.
\newblock \emph{The Annals of Statistics}, 41\penalty0 (4):\penalty0
  2097--2122, 2013.

\bibitem[Bickel and Chen(2009)]{bickel2009nonparametric}
Peter~J Bickel and Aiyou Chen.
\newblock A nonparametric view of network models and newman--girvan and other
  modularities.
\newblock \emph{Proceedings of the National Academy of Sciences}, 106\penalty0
  (50):\penalty0 21068--21073, 2009.

\bibitem[Calderer and Kuijjer(2021)]{calderer2021community}
Gen{\'\i}s Calderer and Marieke~L Kuijjer.
\newblock Community detection in large-scale bipartite biological networks.
\newblock \emph{Frontiers in Genetics}, page 520, 2021.

\bibitem[De~Domenico et~al.(2015)De~Domenico, Nicosia, Arenas, and
  Latora]{de2015structural}
Manlio De~Domenico, Vincenzo Nicosia, Alexandre Arenas, and Vito Latora.
\newblock Structural reducibility of multilayer networks.
\newblock \emph{Nature communications}, 6\penalty0 (1):\penalty0 1--9, 2015.

\bibitem[Flynn and Perry(2020)]{flynn2020profile}
Cheryl Flynn and Patrick Perry.
\newblock Profile likelihood biclustering.
\newblock \emph{Electronic Journal of Statistics}, 14\penalty0 (1):\penalty0
  731--768, 2020.

\bibitem[Harper and Konstan(2015)]{harper2015movielens}
F~Maxwell Harper and Joseph~A Konstan.
\newblock The movielens datasets: History and context.
\newblock \emph{Acm transactions on interactive intelligent systems (tiis)},
  5\penalty0 (4):\penalty0 1--19, 2015.

\bibitem[Ji and Jin(2016)]{ji2016coauthorship}
Pengsheng Ji and Jiashun Jin.
\newblock Coauthorship and citation networks for statisticians.
\newblock \emph{The Annals of Applied Statistics}, 10\penalty0 (4):\penalty0
  1779--1812, 2016.

\bibitem[Jing et~al.(2021)Jing, Li, Lyu, and Xia]{jing2021community}
Bing-Yi Jing, Ting Li, Zhongyuan Lyu, and Dong Xia.
\newblock Community detection on mixture multilayer networks via regularized
  tensor decomposition.
\newblock \emph{The Annals of Statistics}, 49\penalty0 (6):\penalty0
  3181--3205, 2021.

\bibitem[Jing et~al.(2022)Jing, Li, Ying, and Yu]{jing2022community}
Bingyi Jing, Ting Li, Ningchen Ying, and Xianshi Yu.
\newblock Community detection in sparse networks using the symmetrized
  laplacian inverse matrix (slim).
\newblock \emph{Statistica Sinica}, 32\penalty0 (1):\penalty0 1, 2022.

\bibitem[Lancichinetti et~al.(2009)Lancichinetti, Fortunato, and
  Kert{\'e}sz]{lancichinetti2009detecting}
Andrea Lancichinetti, Santo Fortunato, and J{\'a}nos Kert{\'e}sz.
\newblock Detecting the overlapping and hierarchical community structure in
  complex networks.
\newblock \emph{New journal of physics}, 11\penalty0 (3):\penalty0 033015,
  2009.

\bibitem[Li et~al.(2021)Li, Hu, Wang, and Zhang]{li2021super}
Ting Li, Jianchang Hu, Shiying Wang, and Heping Zhang.
\newblock Super-variants identification for brain connectivity.
\newblock \emph{Human brain mapping}, 42\penalty0 (5):\penalty0 1304--1312,
  2021.

\bibitem[Noroozi et~al.(2021{\natexlab{a}})Noroozi, Pensky, and
  Rimal]{noroozi2021sparse}
Majid Noroozi, Marianna Pensky, and Ramchandra Rimal.
\newblock Sparse popularity adjusted stochastic block model.
\newblock \emph{The Journal of Machine Learning Research}, 22\penalty0
  (1):\penalty0 8671--8706, 2021{\natexlab{a}}.

\bibitem[Noroozi et~al.(2021{\natexlab{b}})Noroozi, Rimal, and
  Pensky]{noroozi2021estimation}
Majid Noroozi, Ramchandra Rimal, and Marianna Pensky.
\newblock Estimation and clustering in popularity adjusted block model.
\newblock \emph{Journal of the Royal Statistical Society Series B: Statistical
  Methodology}, 83\penalty0 (2):\penalty0 293--317, 2021{\natexlab{b}}.

\bibitem[Rohe et~al.(2016)Rohe, Qin, and Yu]{rohe2016co}
Karl Rohe, Tai Qin, and Bin Yu.
\newblock Co-clustering directed graphs to discover asymmetries and directional
  communities.
\newblock \emph{Proceedings of the National Academy of Sciences}, 113\penalty0
  (45):\penalty0 12679--12684, 2016.

\bibitem[Sengupta and Chen(2018)]{sengupta2018block}
Srijan Sengupta and Yuguo Chen.
\newblock A block model for node popularity in networks with community
  structure.
\newblock \emph{Journal of the Royal Statistical Society: Series B (Statistical
  Methodology)}, 80\penalty0 (2):\penalty0 365--386, 2018.

\bibitem[Wang et~al.(2023)Wang, Zhang, Liu, Zhu, and Guo]{wang2023fast}
Jiangzhou Wang, Jingfei Zhang, Binghui Liu, Ji~Zhu, and Jianhua Guo.
\newblock Fast network community detection with profile-pseudo likelihood
  methods.
\newblock \emph{Journal of the American Statistical Association}, 118\penalty0
  (542):\penalty0 1359--1372, 2023.

\bibitem[Wang(2010)]{wang2010consistent}
Junhui Wang.
\newblock Consistent selection of the number of clusters via crossvalidation.
\newblock \emph{Biometrika}, 97\penalty0 (4):\penalty0 893--904, 2010.

\bibitem[Wang et~al.(2020)Wang, Liang, and Ji]{wang2020spectral}
Zhe Wang, Yingbin Liang, and Pengsheng Ji.
\newblock Spectral algorithms for community detection in directed networks.
\newblock \emph{The Journal of Machine Learning Research}, 21\penalty0
  (1):\penalty0 6101--6145, 2020.

\bibitem[Wu et~al.(2020)Wu, Zhang, Chen, Guo, and Wang]{wu2020deep}
Ling Wu, Qishan Zhang, Chi-Hua Chen, Kun Guo, and Deqin Wang.
\newblock Deep learning techniques for community detection in social networks.
\newblock \emph{IEEE Access}, 8:\penalty0 96016--96026, 2020.

\bibitem[Zhang et~al.(2021)Zhang, He, and Wang]{zhang2021directed}
Jingnan Zhang, Xin He, and Junhui Wang.
\newblock Directed community detection with network embedding.
\newblock \emph{Journal of the American Statistical Association}, pages 1--11,
  2021.

\bibitem[Zhao et~al.(2012)Zhao, Levina, and Zhu]{zhao2012consistency}
Yunpeng Zhao, Elizaveta Levina, and Ji~Zhu.
\newblock Consistency of community detection in networks under degree-corrected
  stochastic block models.
\newblock \emph{The Annals of Statistics}, 40\penalty0 (4):\penalty0
  2266--2292, 2012.

\bibitem[Zhao et~al.(2024)Zhao, Hao, and Zhu]{zhao2024variational}
Yunpeng Zhao, Ning Hao, and Ji~Zhu.
\newblock Variational estimators of the degree-corrected latent block model for
  bipartite networks.
\newblock \emph{Journal of Machine Learning Research}, 25\penalty0
  (150):\penalty0 1--42, 2024.

\bibitem[Zhou and Amini(2019)]{zhou2019analysis}
Zhixin Zhou and Arash~A Amini.
\newblock Analysis of spectral clustering algorithms for community detection:
  the general bipartite setting.
\newblock \emph{The Journal of Machine Learning Research}, 20\penalty0
  (1):\penalty0 1774--1820, 2019.

\bibitem[Zhou and Amini(2020)]{zhou2020optimal}
Zhixin Zhou and Arash~A Amini.
\newblock Optimal bipartite network clustering.
\newblock \emph{The Journal of Machine Learning Research}, 21\penalty0
  (1):\penalty0 1460--1527, 2020.

\end{thebibliography}

\end{document}